\documentclass[aps,prd,preprint,superscriptaddress,showpacs]{revtex4}
\usepackage{epsfig}
\usepackage{dcolumn}
\usepackage{bm}
\usepackage{amsmath}
\usepackage{amsfonts}
\usepackage{amssymb}
\usepackage{graphicx}
\usepackage{latexsym}
\usepackage{multirow}
\usepackage{amssymb}
\usepackage{amsmath}
\usepackage{amsfonts}
\usepackage{mathrsfs}
\usepackage{ulem}
\usepackage{xcolor}
\usepackage{hyperref}

\def\tl{\tilde}
\def\sl#1{#1\!\!\!\!/}

\begin{document}
\title{Electroweak Supersymmetry (EWSUSY) in the NMSSM}

\author{Taoli Cheng}
\affiliation{State Key Laboratory of Theoretical Physics,
      Institute of Theoretical Physics, Chinese Academy of Sciences,
Beijing 100190, P. R. China}

\author{Tianjun Li}
\affiliation{State Key Laboratory of Theoretical Physics,
      Institute of Theoretical Physics, Chinese Academy of Sciences,
Beijing 100190, P. R. China}
\affiliation{School of Physical Electronics,
University of Electronic Science and Technology of China,
Chengdu 610054, P. R. China}
\affiliation{George P. and Cynthia W. Mitchell Institute for 
Fundamental Physics and Astronomy,
Texas A\&M University, College Station, TX 77843, USA}

\begin{abstract}

To explain all the available experimental results, we have 
proposed the Electroweak Supersymmetry (EWSUSY) previously, 
where the squarks and/or gluino are heavy around a few TeVs 
while the sleptons, sneutrinos, Bino, Winos, and/or Higgsinos 
are light within one TeV. In the Next to Minimal 
Supersymmetric Standard Model (NMSSM), we perform the
systematic $\chi^2$ analyses on parameter space scan for 
three EWSUSY scenarios: (I) $R$-parity conservation and 
one dark matter candidate; (II) $R$-parity conservation 
and multi-component dark matter; (III) $R$-parity violation. 
We obtain the minimal $\chi^2/{\rm (degree ~of ~freedom)}$ 
of 10.2/15, 9.6/14, and 9.2/14 respectively for 
Scenarios I, II, and III. Considering the constraints from 
the LHC neutralino/chargino and slepton searches, we find 
that the majority of viable parameter space prefered 
by the muon anomalous magnetic moment has been excluded 
except for the parameter space with moderate to large 
$\tan \beta$ ($\gtrsim$ 8). Especially, {\it the most 
favorable  parameter space has relatively large $\tan \beta$, 
moderate $\lambda$, small $\mu_{\rm eff}$, heavy squarks/gluino,
and the second lightest CP-even neutral Higgs boson with mass 
around 125 GeV.} In addition, if the left-handed smuon is nearly 
degenerate with or heavier than Wino, there is no definite bound 
on Wino mass. Otherwise, the Wino with mass up to $\sim 450$ GeV 
has been excluded. Furthermore, we present several benchmark 
points for Scenarios I and II, and briefly discuss the prospects 
of the EWSUSY searches at the 14 TeV LHC and ILC. 

\end{abstract}

\maketitle

\newpage
\section{Introduction}

Supersymmetry (SUSY) is the most promising new physics beyond the 
Standard Model (SM). From the theoretical point of view, it solves
the gauge hierarchy problem in the SM, and is consistent with
the Grand Unified Theories (GUTs) due to the gauge coupling unification
in the supersymmetric SMs (SSMs). From the phenomenological
point of view, the electroweak (EW) gauge symmetry can be broken
radiatively due to the large top quark Yukawa coupling, and the lightest
supersymmetric particle (LSP) such as neutralino can be a cold dark matter 
canidiate if $R$-parity is conserved. Also, in the Minimal SSM (MSSM),
the lightest CP-even neutral Higgs boson is predicted to be lighter than
about 130 GeV (For a review, see Ref.~\cite{Djouadi:2005gj}.), 
which is compatible with the Higgs boson with mass
around 125 GeV discovered by the ATLAS and CMS Collaborations 
in July 2012~\cite{Aad:2012tfa, Chatrchyan:2012ufa}.

Inspired by the LHC Higgs~\cite{moriond2013} and SUSY~\cite{LHC-SUSY} searches, 
the experimental results/constraints
on B physics~\cite{Aaij:2012nna, Buchmueller:2009fn} 
and Flavour Changing Neutral Current 
(FCNC)~\cite{Barberio:2007cr, Asner:2010qj, Amhis:2012bh}, anomalous
magnetic momentum of the muon~\cite{Davier:2010nc},  
dark matter relic density from WMAP experiment~\cite{Hinshaw:2012aka}, and 
direct dark matter search from XENON100 experiment~\cite{Aprile:2012nq}, 
we proposed the Electroweak Supersymmetry (EWSUSY):  the squarks and/or gluino 
are heavy around a few TeVs while the sleptons, sneutrinos, Bino, 
Winos, and/or Higgsinos are light within one TeV~\cite{Cheng:2012np}. 
To realize the EWSUSY in the MSSM, we considered 
the non-universal gaugino masses, universal/non-universal scalar masses,
and universal/non-universal trilinear soft terms, inspired by the 
Generalized Minimal Supergravity (GmSUGRA)~\cite{Li:2010xr, Balazs:2010ha}. 
For the later and relevant studies,
see Refs.~\cite{Giudice:2012pf, Ibe:2012qu, Grajek:2013ola,
Endo:2013bba, Mohanty:2013soa,
Bhattacharyya:2013xba, Akula:2013ioa, Iwamoto:2013mya, Choudhury:2013jpa}. 
Moreover, the heavy squarks are prefered by the SUSY FCNC and
CP violation problems. And the light electroweak SUSY 
sector is very promising on both model building and phenomenological study.

Let us briefly review the current LHC Higgs, SUSY and B physics searches
and the anomalous magnetic momentum of the muon in the SSMs.
The ALTAS and CMS Collaborations have released their latest results for the Higgs boson 
searches by various channels~\cite{moriond2013}, which indicate a highly SM-like Higgs
particle with mass around 125 GeV. The concrete experimental results will be given
in Section III. However, the SUSY searches at the LHC still suffer from the null 
results~\cite{LHC-SUSY}. 
Until now,  the first two generation squarks with mass less than around 1.5 TeV have been 
excluded in the Contrainted MSSM (CMSSM) or Minimal Supergravity (mSUGRA).
For the simple decay chains ($\tl g \to t \bar t \tl \chi_1^0$ or 
$\tl g \to b \bar b \tl \chi_1^0$) in the simplified models where all the 
 supersymmetric particles (sparticles) except the LSP neutralino $\chi_1^0$ and gluino
are decoupled,  gluino with mass below about 1.3 TeV has been excluded 
 for the LSP lighter than $\sim 500$ GeV. 
With the LSP below $ 300$ GeV, the masses of light stop and sbottom have been 
pushed up to $\sim 600-700$ GeV in different decay modes except
for the mass-degenerate region. Moreover, the first evidence of 
rare decay $B_s^0 \to \mu^+ \mu^-$ has been found by the LHCb Collaboration
recently~\cite{Aaij:2012nna}. 
The value of branching fraction $3.2^{+1.5}_{-1.2}\times10^{-9}$ leaves 
very small room for the contributions from new physics like SUSY beyond the SM.

Interestingly, the 3.6$\sigma$ deviation of the anomalous magnetic moment of 
the muon $(g_{\mu}-2)/2$: 
$\Delta a_{\mu}=a_{\mu}^{\rm exp}-a_{\mu}^{\rm SM}=(2.87\pm0.8)\times10^{-9}$~\cite{Davier:2010nc} 
may imply the new physics beyond the SM around the electroweak scale. 
In the SSMs, the light smuon, muon-sneutrino, Bino, Winos, and Higgsinos would contribute 
to $\Delta a_{\mu}$~\cite{Moroi:1995yh,Martin:2001st,Byrne:2002cw,Stockinger:2006zn,Domingo:2008bb}. 
The SUSY contributions could roughly be approximated as 
$\sim 1.3 \times 10 ^{-9}{\rm sgn}(\mu M_{2})\left(\frac{100 {\rm GeV}}{M_{\rm SUSY}}\right)^2 \tan \beta$, 
where $M_{\rm SUSY}$ denotes the typical mass scale of relevant sparticles. 
With $M_{\rm SUSY} \sim$ several hundred GeV and $\tan \beta \sim \mathcal{O}(10)$, the discrepancy 
can be explained.
However, to obtain the Higgs boson mass 125 GeV in the CMSSM/mSUGRA,
we require relatively large gaugino mass $M_{1/2}$ and/or scalar mass $M_0$,
 which correlate the squark and slepton masses. Thus, it is very
difficut to obtain the above $\Delta a_{\mu}$ in the CMSSM/mSUGRA. Of course,
the above $\Delta a_{\mu}$ can be realized in the MSSM with the EWSUSY due to the 
non-universal gaugino masses, scalar masses, as well as trlinear soft terms~\cite{Cheng:2012np}.

The simplest extension of the MSSM is the Next-to-MSSM (NMSSM), 
where a SM singlet Higgs field $S$ is introduced (For a review, see
Ref.~\cite{Ellwanger:2009dp}.). In the NMSSM,
we can solve the $\mu$ problem dynamically while keep the 
above attractive features 
in the MSSM. In fact, there exists some degree of fine-tuning
to have the Higgs boson mass about 125 GeV in the MSSM.
Interestingly, we can lift the SM-like Higgs boson mass and then
solve such fine-tunging problem in the NMSSM for two reasons:
(1) the F-term contribution to the tree-level Higgs potential
from superpotential term $\lambda S H_d H_u$, where $H_d$ and $H_u$
are one pair of Higgs doublets in the SSMs. This contribution to
the Higgs mass square is proportional to $\lambda^2 \sin^22\beta$, where
 $\tan\beta$ is the ratio of the Vacuum Expectation Values (VEVs)
of $H_u$ and $H_d$; (2) The pushing up effect from the diagonalization
of Higgs boson mass matrix since the second lightest CP-even neutral
Higgs boson can be SM-like. If the Higgs boson decay branching 
fractions were indeed deviated from the SM predictions, such devivations may 
be explained in the NMSSM as well. Thus, the NMSSM has been studied extensively
after the Higgs boson discovery~\cite{Ellwanger:2011aa, Gunion:2012zd,
King:2012is, Kang:2012sy, Ellwanger:2012ke, Gunion:2012gc, Kowalska:2012gs,
NMSSM-REFs, Badziak:2013bda}. To obtain the SM-like Higgs boson mass
around 125 GeV in the previous studies of the constrained NMSSM, one 
usually considered the small $\tan\beta$ ($\tan\beta \le 4$)  and 
large $\lambda$ ($\lambda \sim 0.6-0.7$) regime, which gives relatively large 
tree-level F-term contribution. However, the SUSY contributions to
the anomalous magnetic moment of 
the muon $\Delta a_{\mu}$ are generically smaller
than about $4.0\times 10^{-10}$~\cite{Gunion:2012zd, Kowalska:2012gs}.

In this paper, with the GmSUGRA for supersymmtry breaking soft terms,
we consider the following three EWSUSY scenarios in the NMSSM: 
(I) $R$-Parity Conservation (RPC), and the LSP neutralino is the only
 dark matter candidate;  (II) $R$-parity conservation and 
multi-component dark matter. So the 
LSP neutralino relic density just needs to be smaller than the observed 
dark matter density; (III) $R$-Parity Violation (RPV). So
the LSP can be the lightest neutralino, light stau, or tau sneutrino,
which  will decay into the SM particles through RPV superpotential terms. 
In the parameter space scan, we consider the experimental results from 
the LHC Higgs seaches, B physics,  $\Delta a_{\mu}$, and dark matter
relic density, etc. The Degree Of Freedom (DOF) from 
the experimental data is
15, 14, and 14 respectively for Scenarios I, II, and III.
We perform the detailed $\chi^2$  analyses,
and find that the minimal $\chi^2/{\rm DOF}$ 
are 10.2/15, 9.6/14, and 9.2/14 respectively for Scenarios I, II, and III.
Similar to the previous studies, we still obtain some viable parameter 
space where the light stop and gluino are light within 1 TeV
for small $\tan\beta \le 4$ and large $\lambda\sim 0.6-0.7$.
Employing the constraints from the LHC neutralino/chargino 
and slepton searches, we show that the majority of viable 
parameter space prefered by $\Delta a_{\mu}$
has been excluded except for the parameter space with moderate 
to large $\tan \beta$ ($\gtrsim$ 8). In particular, {\it in
the most favorable  parameter space, we have relatively large 
$\tan \beta$, moderate $\lambda$,  small $\mu_{\rm eff}$,
heavy squarks/gluino, and the second lightest CP-even neutral 
Higgs boson being SM-like.} The SM-like Higgs boson mass is
lifted by the radiative corrections due to the large stop masses 
and pushing up effect~\cite{Kang:2012sy}. 
In addition, if the left-handed smuon is nearly 
degenerate with or heavier than Wino, there is no definite bound
on Wino mass, which can be as light as 230~GeV. While if the 
left-handed smuon is lighter than Wino, the Wino with mass 
below $\sim 450$ GeV has been excluded. Furthermore, we present
a few benchmark points for Scenarios I and II where $\chi^2/{\rm DOF}$
can be around one, and we briefly discuss the prospects 
of the EWSUSY searches at the $\sqrt{s}=14$~TeV LHC (LHC-14) and ILC
($e^+ e^-$ linear collider with designed center mass energy of $500$ GeV -- 1 TeV).

This paper is organized as follows. In Section II, we give a brief review of the EWSUSY
from the GmSUGRA and the supersymmetric contributions to the anomalous magnetic moment of the muon. 
In Section III, the numerical results employing $\chi^2$ statistic test are given. 
And we present the systematic analyses of the resulting distribution of viable parameter
space and sparticle spectra. 
In Section IV, the LHC neutralino/chargino and slepton search constraints are considered, and 
the  future searches of the electroweak SUSY sector are discussed. We also present 
some benchmark points. Finally in Section V, we summarize our work briefly.

\section{A Brief Review of the EWSUSY and $(g_{\mu}-2)/2$ }

In this Section, we will briefly review the EWSUSY from the GmSUGRA as well as 
the SUSY contributions to the anomalous magnetic moment of the muon.

\subsection{The EWSUSY from the GmSUGRA in the NMSSM}

The EWSUSY, which can be realized in the GmSUGRA~\cite{Li:2010xr},  
has a mass hierarchy for the colored and uncolored sparticles: 
squarks and/or gluino are heavy around several TeV while the sleptons, sneutrinos, Bino, Winos, and/or Higgsinos 
 are light within one TeV~\cite{Cheng:2012np}. 
In the GmSUGRA, the gauge coupling relation and gaugino mass relation at the GUT scale are
\begin{equation}
 \frac{1}{\alpha_2}-\frac{1}{\alpha_3} =
 k~\left(\frac{1}{\alpha_1} - \frac{1}{\alpha_3}\right)~,
\end{equation}
\begin{equation}
 \frac{M_2}{\alpha_2}-\frac{M_3}{\alpha_3} =
 k~\left(\frac{M_1}{\alpha_1} - \frac{M_3}{\alpha_3}\right)~,
\end{equation}
where $k$ is the index and equal to 5/3 in the simple GmSUGRA. Assuming a univeral gauge coupling 
at the GUT scale ($\alpha_1=\alpha_2=\alpha_3$) for simplicity, we obtain a simple gaugino mass relation
\begin{equation}
 M_2-M_3 = \frac{5}{3}~(M_1-M_3)~.
\end{equation}
The univeral gaugino mass relation in the mSUGRA $M_1 = M_2 = M_3$ is just a special case of 
this general one. Choosing $M_1$ and $M_2$ to be free input parameters which vary around 
several hundred GeV for the EWSUSY, we get   
\begin{eqnarray}
M_3=\frac{5}{2}~M_1-\frac{3}{2}~M_2~,
\end{eqnarray}
which could be as large as several TeV or as small as several hundred GeV, depending
 on specific values of $M_1$ and $M_2$.

The general supersymmetry breaking scalar masses at the GUT scale are given 
in Ref.~\cite{Balazs:2010ha}. 
Taking the slepton masses  to be free, we obtain the following squark masses 
\begin{eqnarray}
M_{\tl{Q}_i}^2 &=& \frac{5}{6} (M_0^{U})^2 +  \frac{1}{6} M_{\tl{E}_i^c}^2~,\\
M_{\tl{U}_i^c}^2 &=& \frac{5}{3}(M_0^{U})^2 -\frac{2}{3} M_{\tl{E}_i^c}^2~,\\
M_{\tl{D}_i^c}^2 &=& \frac{5}{3}(M_0^{U})^2 -\frac{2}{3} M_{\tl{L}_i}^2~,
\end{eqnarray}
where $M_{\tl Q}$, $M_{\tl U^c}$, $M_{\tl D^c}$, $M_{\tl L}$, and  $M_{\tl E^c}$ denote the scalar masses of
the left-handed squark doublets, right-handed up-type squarks, right-handed down-type squarks,
left-handed sleptons, and right-handed sleptons, respectively. Also, $M_0^U$ is the univeral  
scalar mass, as in the mSUGRA. In the EWSUSY, $M_{\tl L}$ and $M_{\tl E^c}$ are both within 1 TeV, resulting in 
light sleptons. Especially, in the limit $M_0^U \gg M_{\tl L/\tl E^c}$, we have the approximated 
relations for squark masses: $2 M_{\tl Q}^2 \sim M_{\tl U^c}^2 \sim M_{\tl D^c}^2$. In addition, 
the Higgs soft masses $M_{\tl H_u}$ and $M_{\tl H_d}$, and the  trilinear soft terms
 $A_U$, $A_D$ and $A_E$ can all be free parameters from the GmSUGRA~\cite{Cheng:2012np, Balazs:2010ha}.

With the SM-like Higgs boson mass around 125 GeV, we can still have much lighter stops and gluino in the
NMSSM than the MSSM, as explained in the Introduction.
Thus, different from our previous EWSUSY work in the MSSM~\cite{Cheng:2012np}, 
in general,  the squarks and gluino in the NMSSM 
can be either light below 1 TeV, or heavy about several TeV, beyond the reach of the current LHC. 
Let us classify the
different combinations of $M_3$ and $M_0$, which would produce the very characteristic 
particle spectra of squarks and gluino in the following:
\begin{itemize}
 \item Large $M_3$ and large $M_0$: all colored sparticles are definitely decoupled with masses around several TeV. 
 \item Large $M_3$ and small $M_0$: similar to the case above. 
 \item Small $M_3$ and large $M_0$: gluino is light around or below 1 TeV, while squarks are heavy. 
The mass squares of  right-handed squarks are predicted to be approximately 
twice those of left-handed ones. 
 \item Small $M_3$ and small $M_0$: all the colored sparticles could be light about $\sim 1-2$ TeV. Because the light 
first two generation squarks are disfavored by the null results of the LHC SUSY searches, this case could 
survive only with the special sparticle spectra such as compressed ones or with the RPV.

\end{itemize}

The superpotential and SUSY breaking soft terms for the Higgs sector in the NMSSM are
\begin{equation}
 W = \lambda S H_u H_d + \frac{\kappa}{3} S^3~,
\end{equation}
\begin{equation}
 V_{\rm soft}=m_{H_u}^2|H_u|^2 + m_{H_d}^2|H_d|^2+m_S^2|S|^2+
 \left(\lambda A_{\lambda} S H_u H_d + \frac{1}{3} \kappa A_{\kappa}
 S^3 + h.c \right)~.
\end{equation}
After spontaneous electroweak symmetry breaking, the singlet scalar Higgs field $S$ obtains a VEV, and then
 the effective $\mu$ term is generated dynamically, {\it i.e.}, $ \mu_{\rm eff} = \lambda \langle S \rangle$. 
Also, in the decoupling limit and providing that there is no mixing between the Higgs doublets and singlet,
we have the following tree-level Higgs boson mass 
\begin{equation}
 M_H^2 = m_Z^2 \cos^2 2\beta + \lambda^2 v^2 \sin^2 2\beta~, 
\end{equation}
where $v^2 = (\langle H_u^0 \rangle)^2+(\langle H_d^0 \rangle)^2=(174 {\rm GeV})^2$.
Comparing to the Higgs sector in the MSSM, we have an extra
 SM singlet Higgs field $S$ and the mixings between Higgs doublets and singlet.
 To obtain a 125 GeV SM-like Higgs particle, one usually considered 
the small $\tan \beta \sim 2$ and large $\lambda$ $\sim 0.6$ in the previous studies so that 
the new tree-level contribution can be large.  Moreover, the
proper doublet-singlet mixing can shift the Higgs boson mass by several GeV. 
As we know, the SM-like Higgs boson in the NMSSM can be either the lightest ($H_1$)
or the second lightest ($H_2$) CP-even neutral Higgs boson. For
the latter case it is more efficient to lift the SM-like Higgs boson mass up to $\sim 125$ GeV while 
obtain the enhancement in di-photon channel through the suitable H/S mixing previously. 
To be concrete, with a lighter $H_1$ being singlet-like, the SM-like Higgs boson $H_2$ would 
be pushed up by the diagonalization of Higgs boson mass
 matrix~\cite{Kang:2012sy}. 
Thus,  a 125 GeV SM-like Higgs boson can be realized in the NMSSM  
without very heavy stops and gluino.

\begin{figure}[htb!]
 \begin{center}
  \includegraphics[width=0.7\textwidth]{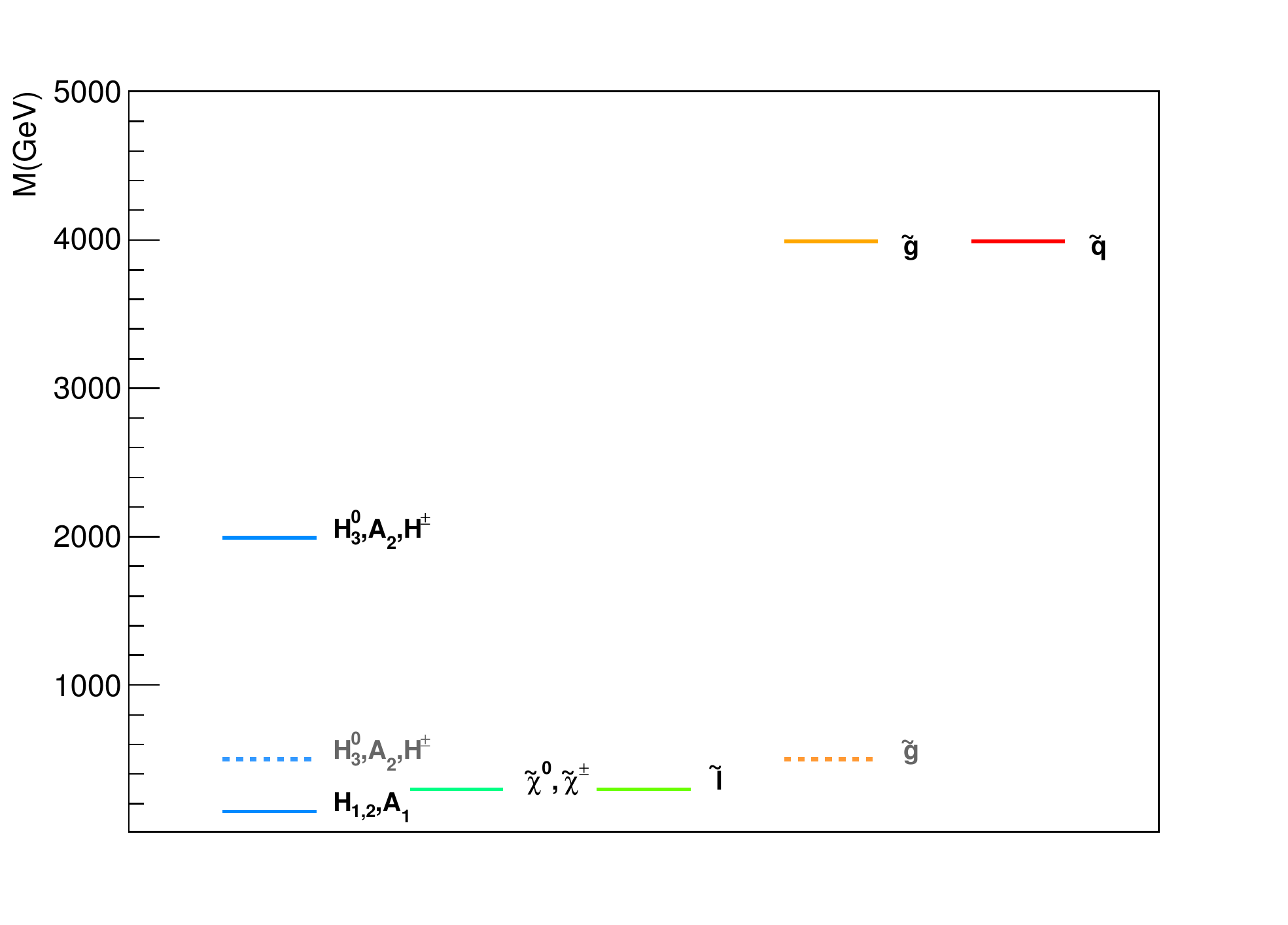}
  \caption{\label{spectrum} The typical EWSUSY mass hierarchy for the particle spectra in the NMSSM. The
dashed lines denote alternative cases.}
 \end{center}
\end{figure}

We display the characteristic mass hierarchy for the EWSUSY in the NMSSM in Fig.~\ref{spectrum},
where the dashed lines denote alternative cases. The squarks are heavy around several TeV, and 
gluino can be either heavy or light, although the light gluino is 
strongly constrained by the recent LHC SUSY searches. All the neutralinos, charginos, and
sleptons, are light around several hundred GeV. The Higgsinos, which can be heavy, are also  around electroweak 
scale, as the results of small $\mu_{\rm eff}  \sim 100-300$ GeV from the following scan 
in the NMSSM. The features of Higgs sector are inspired by general NMSSM properties. 
At least two CP-even ($H_1$ and $H_2$) and one CP-odd ($A_1$) neutral Higgs fields are light around $ 100$ GeV, being 
$H_u$-like or S-like. The other heavy Higgs fields $H_3$, $A_2$ and $H^{\pm}$ being $H_d$-like can
 be either light around $ 500$ GeV or heavy about several TeV.  
All these features will be explained in details in the following Section. 

\subsection{SUSY Contributions to $(g_{\mu}-2)/2$}

The dominant SUSY contributions to the muon anomalous magnetic moment
$\Delta a_\mu$ arise from the neutralino-smuon and chargino-sneutrino loops. 
And these contributions in the NMSSM are similar to the MSSM.  
In the NMSSM, the extra contribution would come from a very light pesdo-scalar 
Higgs boson ($\sim$ several GeV)~\cite{Domingo:2008bb}, which
 can be neglected in the following discussions.

The contributions to $\Delta a_\mu$ from the neutralino-smuon and chargino-sneutrino loops are~\cite{Martin:2001st}
\begin{align}
 \Delta a_{\mu}^{\tilde \chi^0}&=\frac{m_\mu}{16\pi^2}\sum_{i,m}\left\{-\frac{m_\mu}{12m_{\tilde \mu_m}^2}(|n_{im}^L|^2+|n_{im}^R|^2) F_1^N(x_{im})+\frac{m_{\tilde \chi_i^0}}{3m_{\tilde \mu_m}^2}{\rm Re}[n_{im}^L n_{im}^R]F_2^N(x_{im})\right\} \label{amu_neu}~,~\\
  \Delta a_{\mu}^{\tilde \chi^\pm}&=\frac{m_\mu}{16\pi^2}\sum_{k}\left\{\frac{m_\mu}{12m_{\tilde \nu_\mu}^2}(|c_{k}^L|^2+|c_{k}^R|^2) F_1^C(x_{k})+\frac{2m_{\tilde \chi_k^\pm}}{3m_{\tilde \nu_\mu}^2}{\rm Re}[c_k^L c_k^R]F_2^C(x_k)\right\} \label{amu_cha} ~,~
\end{align}
where $i=1,2,3,4,5$, $k=1,2$, and $m=1,2$ denote the mass eigenstates of neutralinos, charginos, and smuons, respectively.
The kinematic variables are $x_{im}=m_{\tilde \chi_i^0}^2/m_{\tilde \mu_m}^2$, and $x_k=m_{\tilde \chi_k^\pm}^2/m_{\tilde \nu_\mu}^2$. 
The couplings are given by
\begin{eqnarray}
 n_{im}^L&=&~\frac{1}{\sqrt{2}}(g_2 N_{i2} + g_1 N_{i1})X_{m1}^* - y_\mu N_{i3} X_{m2}^*~, \\
 n_{im}^R&=&~\sqrt{2} g_1 N_{i1} X_{m2} + y_\mu N_{i3} X_{m1}~, \\
 c_k^L &=&~ -g_2 V_{k1}~,\\
 c_k^R &=&~ y_\mu U_{k2}~,
\end{eqnarray}
where $X_{m-}$, $N_{i-}$, and $U_{k-}$/$V_{k-}$ are the elements of the conventional 
mixing matrices for smuons, neutralinos, and charginos, respectively. 
In the gauge eigenstate bases, $\tilde \chi^0(G)=(-i\tilde B, -i\tilde W, \tl H^0_d, \tl H^0_u, \tl S)$, 
the $i$-th neutralino $ \tl \chi_j^0$ mass eigenstate is equal to $N_{ij} \tl \chi_j^0(G)$. 
Similarly, we have $\tl \chi_k^+=V_{kl} \tl \chi_l^+(G)$, $\tl \chi_k^-=U_{kl} \tl \chi_l^-(G)$,
 and $\tl \mu_m=X_{mn} \tl \mu_n(G)$ in the gauge eigenstate bases
 $\tl \chi^+(G)=(-i \tl W, \tl H_u^+)$, $\tl \chi^-(G)=(-i \tl W, \tl H_d^-)$,
 and $\tl \mu(G)=(\tl \mu_L, \tl \mu_R)$, respectively.
Also, $y_\mu= m_\mu/(v \cos \beta$) ($\sim m_\mu \tan \beta/v$ for large $\tan \beta$) 
is the muon Yukawa coupling. Loop functions $F_{1/2}^{N/C}(x)$ are normalized to 1 for $x=1$. 
The concrete mixing matrices and loop functions were given in 
Refs.~\cite{Moroi:1995yh,Martin:2001st,Byrne:2002cw,Stockinger:2006zn}. 

Because the magnetic moment operator is a chirality-flipping interaction, it is
 proportional to $m_{\mu}$ for external-leg chirality flipping (the first term in Eqs.~(\ref{amu_neu}) 
and (\ref{amu_cha})), and to Yukawa coupling for the internal-line chirality flipping 
(the second term in Eqs.~(\ref{amu_neu}) and (\ref{amu_cha})). The internal-line chirality-flipping 
terms would dominate since sparticles are much heavier than muon. 
The contributions from the neutralino-smuon and chargino-sneutrino loops can approximately be expressed as
\begin{align}
 \Delta a_{\mu}^{\tilde \chi^0 \tilde \mu} &\simeq \frac{1}{192\pi^2} \frac{m_{\mu}^2}{M_{SUSY}^2} \left({\rm sgn(\mu M_1)}g_1^2-{\rm sgn(\mu M_2)}g_2^2\right)\tan \beta~, \\
 \Delta a_{\mu}^{\tilde \chi^{\pm} \tilde \nu} &\simeq {\rm sgn(\mu M_2)}\frac{1}{32\pi^2} \frac{m_{\mu}^2}{M_{SUSY}^2} g_2^2 \tan \beta~.
 \end{align}
 Obviously, if all the relevant sparticles are at the same mass scale, the chargino-sneutrino loop 
contributions would dominate. So we have 
$\Delta a_{\mu} \sim 10^{-9} \left(\frac{100~{\rm GeV}}{M_{\rm SUSY}}\right)^2 \tan \beta$ for 
sgn$(\mu M_2)>0$. It was found in Ref.~\cite{Byrne:2002cw} that the 2$\sigma$ bound on $\Delta a_{\mu}$ can 
be achieved for $\tan \beta=10$ if four relevant 
sparticles are lighter than $600-700$ GeV.
While for smaller $\tan \beta$ ($\sim$3), the lighter sparticles ($\lesssim 500$ GeV) are needed. 
All the relevant sparticles can be as light as several hundred GeV 
in the NMSSM with the EWSUSY. Thus, it would be 
very promising to explain the measured  $\Delta a_\mu$ deviation. 


\section{Numerical Analyses}

\subsection{Setup}

In this subsection, we will use $\chi^2$ statistic test to explore the EWSUSY parameter space in the NMSSM. 
We will take $\kappa$ and $\mu_{\rm eff}$ at $M_{\rm SUSY}$ as input parameters instead of the Higgs soft masses 
$M_{H_u}$ and $M_{H_d}$ at the GUT scale. So we have 13 free parameters in total as follows
\begin{center}
 $M_{\tilde L}$, $M_{\tilde E}$, $M_1$, $M_2$, $M_0$, $A_0$, $A_E$, $A_{\lambda}$, $A_{\kappa}$, $\lambda$, $\kappa$, $\tan \beta$, $\mu_{\rm eff}$~,~\,
\end{center}
where we assume $A_U=A_D=A_0$.
Among these input parameters, $\lambda$, $\kappa$, $\tan \beta$, $\mu_{\rm eff}$, $A_{\lambda}$, $A_{\kappa}$, and $A_0$ 
are relevant to Higgs sector. $M_{\tilde L}$, $M_{\tilde E}$, $\mu_{\rm eff}$, $M_1$, and $M_2$ take control of the electroweak SUSY 
sector, and are set to be light around several hundred GeV. The input parameter ranges employed in numerical scan are 
the following: $M_0 \in (0, 3000)$ GeV, $M_1 \in (-1000, 1000)$ GeV, $M_2 \in (-1000,1000)$ GeV, $M_{\tilde L} \in (0, 800)$ GeV, $M_{\tilde E} \in (0, 800)$ GeV, $\lambda \in (0, 0.7)$, $\tan \beta \in (1, 60)$. All the other parameters such as 
$A_0$ are just left free.


As explained in the Introduction, we consider three EWSUSY scenarios in the NMSSM. 
In Scenario I, the LSP neutralino is the only dark matter candidate, and then
the dark matter relic density from WMAP experiment is involed in $\chi^2$ analyses. 
In Scenario II, we assume the multi-component dark matter, and then there is  
an upper bound on the LSP neutralino relic density ($\Omega h^2 < 0.136$). 
Considering the variations in the local relic abundance, we rescale the WIMP-proton scattering cross section
$\sigma^{SI}_p$ to $\sigma^{SI}_p\times \Omega h^2/0.11$ so that 
the XENON100 upper limit~\cite{Aprile:2012nq} can be applied directly. The
previous SUSY search constraints from the LEP and Tevatron are considered in Scenarios I and II as well.
In Scenario III, $R$-parity is violated, and then there is no stable LSP. In general, the RPV couplings are too 
small to shift the sparticle masses much~\cite{Allanach:2006st}. Thus, we just ignore the RPV couplings 
in Renormalization Group Equation (RGE) running. 
In this scenario, we relax the LEP SUSY search constraints,
 and the requirement of the ``lightestness'' of the lightest neutralino, which means that 
any sparticle can be the LSP. There exist constraints for the RPV SUSY from the LEP2~\cite{rpvboundslep} 
and LHC~\cite{LHC-SUSY}. However, 
these existing bounds are highly model dependent. So we just neglect them 
in the following.

Taking into account the uncertainties for theoretical Higgs boson mass calculations, we require
the SM-like Higgs boson mass to be within the range $125.5 \pm 1.5$ GeV. The SM-like Higgs boson at the LHC
 can be the lightest CP-even neutral Higgs boson $H_1$, or the second lightest CP-even one $H_2$, or even both of them.
For the last case in which the two light CP-even neutral Higgs bosons are highly mass degenerate (say,
$\Delta m <2$ GeV), we combine these two Higgs boson productions and decays as Ref.~\cite{Gunion:2012gc}
did to form one Higgs particle observed at the LHC. The effective mass and signal strength for decay channel XX are
defined as follows
\begin{align}
&  m_h^{XX}\equiv \frac{R^{XX}_1 m_1+R^{XX}_2m_2}{R^{XX}_1+R^{XX}_2}~, \\
&  R_h^{XX} \equiv R^{XX}_1 + R^{XX}_2 ~,
\end{align}
where the Higgs signal strength is defined as
\begin{equation}
 R_{XX}=\frac{\sigma(pp \to H)}{\sigma(pp \to H_{SM})} \times
 \frac{BR(H \to XX)}{BR(H_{SM} \to XX)}~.
\end{equation}

In the numerical study, the statistic test $\chi^2$ is constructed in the following simple form
\begin{equation}
 \chi^2= \sum_i \frac{(x_i-x_i^0)^2}{\sigma_i^2}~,
\end{equation}
where $x_i^0$ and $\sigma_i$ are respectively the experimental central value and error of 
the $i$-th observable, while $x_i$ denotes the model prediction. We consider the following 
experimental results for  $\chi^2$ in the code: the Higgs signal strengths of main search channels from 
the ATLAS and CMS Collaborations, low energy phenomenological constraints including B physics, 
anomalous magnetic moment of the muon, and dart matter relic density (for Scenario I only). 
All these constraints are listed in Table~\ref{lhchig}. Thus, the degree of freedom  
is 15 for Scenario I and 14 for Scenarios II and III.

\begin{table}[htb!]
 \begin{center}
  \begin{tabular}{|c c|c|c|} \hline
        & 25 $\rm {fb^{-1}}$& $\gamma \gamma$ & $1.65 \pm 0.32$ \\
        & 25 $\rm {fb^{-1}}$& ZZ & $1.7 \pm 0.45$ \\
   ATLAS$^{7+8}$ & 25$\rm {fb^{-1}}$& WW & $1.01 \pm 0.31$ \\
        & 18$\rm {fb^{-1}}$& Vbb & $-0.4 \pm 1.0$ \\
        & 18$\rm {fb^{-1}}$& $\tau \tau$ & $0.7 \pm 0.7$ \\ \hline
        & 25 $\rm {fb^{-1}}$& $\gamma \gamma$ & $1.11 \pm 0.31$(Cut-based) \\
        & 25 $\rm {fb^{-1}}$& ZZ & $0.91 \pm 0.27$ \\
   CMS$^{7+8}$  & 24 $\rm {fb^{-1}}$& WW & $0.76 \pm 0.21$ \\
       & 17 $\rm {fb^{-1}}$& Vbb & $1.3 \pm 0.65$ \\
       & 24 $\rm {fb^{-1}}$& $\tau \tau$ & $1.1 \pm 0.4$ \\ \hline
   \multicolumn{3}{|c|}{${\rm BR}(B\rightarrow X_s\gamma)$}&$(3.55\pm0.256)~~\times 10^{-4}$~\cite{Amhis:2012bh} \\ 
   \multicolumn{3}{|c|}{${\rm BR}(B_{s}^{0}\rightarrow\mu^{+}\mu^{-})$}&
$3.2^{+1.5}_{-1.2}\times10^{-9}$~\cite{Aaij:2012nna} \\ 
   \multicolumn{3}{|c|}{${\rm BR}(B \rightarrow \tau \nu_\tau)$}&$(1.67\pm0.3)~~\times 10^{-4}$~\cite{Amhis:2012bh} \\ 
   \multicolumn{3}{|c|}{$\Delta a_{\mu}$} & $(2.87\pm0.8)~~\times 10^{-9}$~\cite{Davier:2010nc} \\ 
   \multicolumn{3}{|c|}{$\Omega h^2$} & $0.1157\pm 0.0023$~\cite{Hinshaw:2012aka} \\ \hline
  \end{tabular}
  \caption{\label{lhchig} The LHC Higgs signal strengths~\cite{moriond2013} after Moriond 2013,
and the other experimental constraints.}
 \end{center}
\end{table}

In our numerical study, the NMSSMTools~\cite{Ellwanger:2004xm} is used for the RGE runnig and 
the calculations of low energy phenomenological constraints. 
We have adapted the source code to accomodate the general SUSY breaking soft terms in the GmSUGRA.
The LSP neutralino relic density  
and LSP neutralino-proton spin-independent scattering cross section are calculated via
 the micrOMEGAs~\cite{Belanger:2004yn} implemented in the NMSSMTools package.
 For simplicity, we take the top quark pole mass as 173.5 GeV. To be more efficient in computation, we
use  the Markov Chain Monte Carlo (MCMC) technique for parameter space scan.

\subsection{Numerical Results}

\begin{figure}[ht]
\begin{center}
\includegraphics[width=0.9\textwidth]
{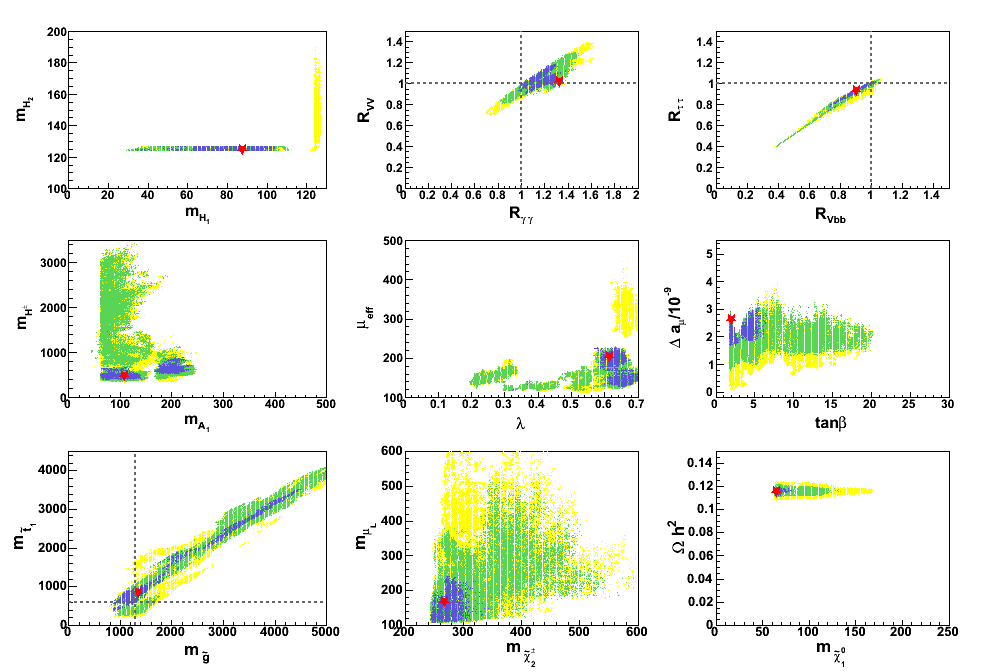}  
\end{center}
\caption{\label{s1} Summary results for the parameter space scan in Scenario I with $\chi^2_{min}/{\rm DOF}=10.2/15$. 
In all the panels, the best-fit point is marked with red pentagram. 1$\sigma$, 2$\sigma$, and
3$\sigma$ regions are colored in purple, green, and yellow, respectively. All the particle masses
and $\mu_{\rm eff}$ are displayed 
in unit GeV. The naive bounds from SUSY searches at the LHC are also shown in $m_{\tl g}-m_{\tl t_1}$ panel, with 
 1.3 TeV for gluino and  600 GeV for light stop.}
\end{figure}

\begin{figure}[ht]
\begin{center}
\includegraphics[width=0.9\textwidth]
{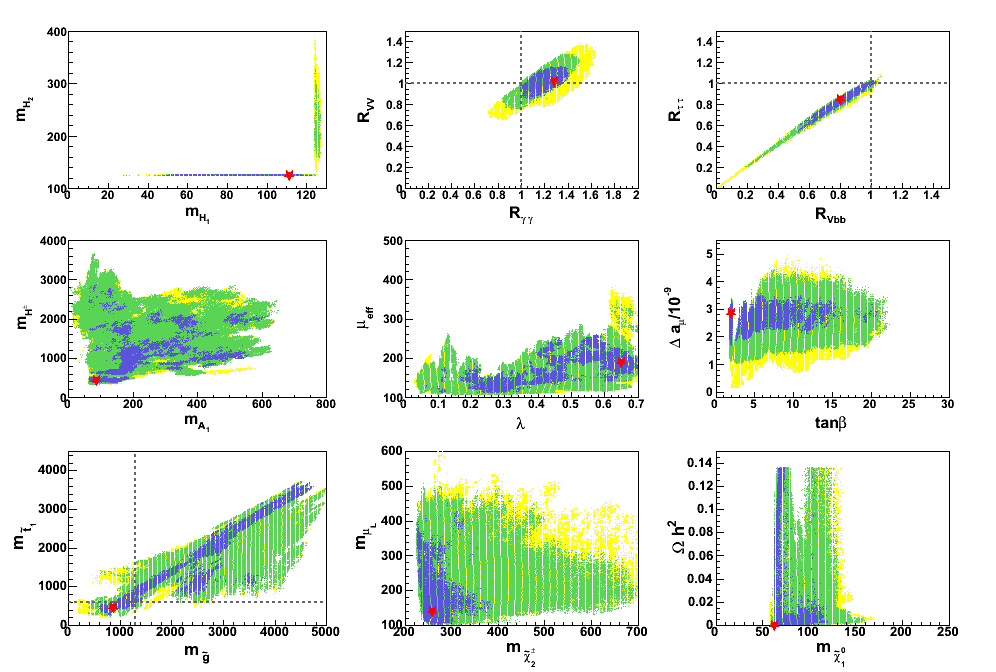}
\end{center}
\caption{\label{s2}  Summary results for the parameter space scan in Scenario II with $\chi^2_{min}/{\rm DOF}=9.6/14$. 
In all the panels, the best-fit point is marked with red pentagram. 1$\sigma$, 2$\sigma$, and
3$\sigma$ regions are colored in purple, green, and yellow, respectively. All the particle masses 
and $\mu_{\rm eff}$ are displayed 
in unit GeV. The naive bounds from SUSY searches at the LHC are also shown in $m_{\tl g}-m_{\tl t_1}$ panel, with 
 1.3 TeV for gluino and  600 GeV for light stop.}
\end{figure}

\begin{figure}[ht]
\begin{center}
\includegraphics[width=0.9\textwidth]
{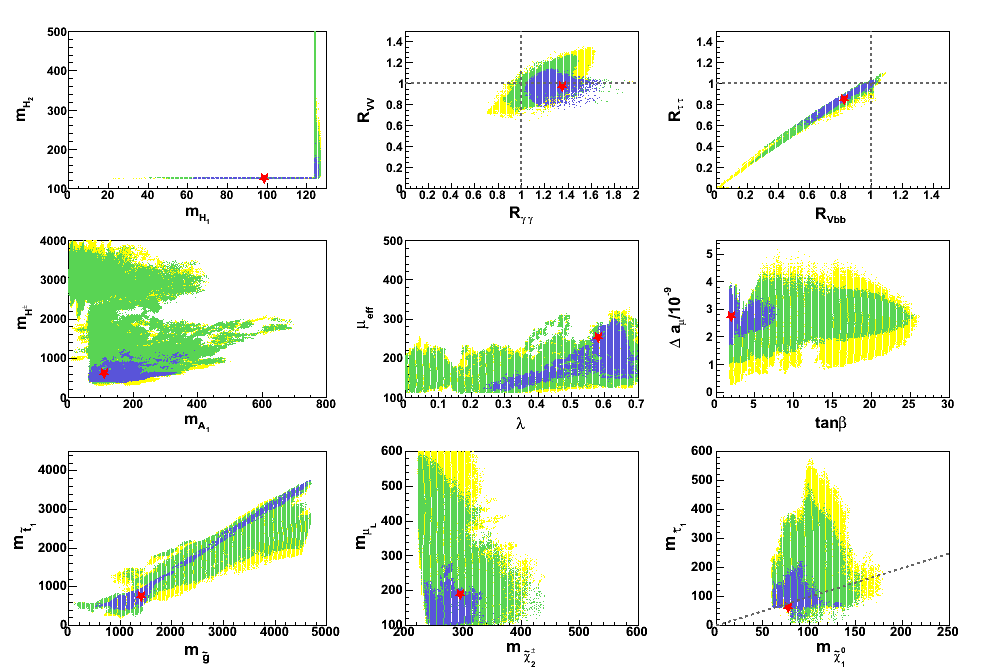}
\end{center}
\caption{\label{s3} Summary results for the parameter space scan in Scenario III with $\chi^2_{min}/{\rm DOF}=9.2/14$. 
In all the panels, the best-fit point is marked with red pentagram. 1$\sigma$, 2$\sigma$, and
3$\sigma$ regions are colored in purple, green, and yellow, respectively. All the particle masses 
and $\mu_{\rm eff}$ are displayed 
in unit GeV. Dashed line in $m_{\tl \chi_1^0}-m_{\tl \tau_1}$ panel is the seperation line with $m_{\tl \chi_1^0}=m_{\tl \tau_1}$.}
\end{figure}

We scan the whole parameter space systematically,
and obtain the minimal $\chi^2/{\rm DOF}$ 10.2/15, 
 9.6/14, and 9.2/14 respectively in Scenarios I, II, and III. We list the three best-fit points in Table~\ref{bptable}. 
The summarized results are shown in two-dimensional (2D) panels in Figs. \ref{s1}, \ref{s2}, and \ref{s3} 
for Scenarios I, II, and III, respectively. We display the 
$1\sigma$~($\Delta \chi^2=\chi^2 -\chi_{\rm min}^2 <2.3$), $2 \sigma$~($\Delta \chi^2<6.2$), 
and $3\sigma$~($\Delta \chi^2<11.8$) regions in every 2D distribution with the other parameters marginalized.
For all three scenarios, we discuss the NMSSM input parameters, 
the Higgs sector, and the  most relevant sparticle masses. Moreover, 
we explain the LSP neutralino in details for Scenarios I and II, and
comment on the possible LSP for Scenario III.

\begin{itemize}

\item The NMSSM Input Parameters and $\Delta a_{\mu}-\tan \beta$
 
From the $\lambda-\mu_{\rm eff}$ and $\Delta a_{\mu}-\tan \beta$ plots in Figs. \ref{s1}, \ref{s2}, and \ref{s3},
we find that small $\mu_{\rm eff}$ ($\lesssim 300~\rm {GeV}$) are prefered in all three scenarios. 
In Scenario I, small $\tan \beta$ and large $\lambda$ are favored as expected from the  
1$\sigma$ region of $\chi^2$ annalyses. The SM-like Higgs boson is the second lightest CP-even neutral
Higgs boson $H_2$, whose mass can be lifted by the large 
tree-level contribution and pushing up effect. Thus, the 
light stop and gluino can indeed be light $\lesssim 1~\rm {TeV}$ in this regime. Previously,
this case has been studied extensively.
However, in Scenarios II and III which relax the dark matter relic abundance requirement, 
we obtain the other interesting parameter space: 1$\sigma$ region of $\chi^2$ analyses covers the wider ranges 
of $\tan \beta$ and $\lambda$, for example,  $\tan \beta$ from 2 to 20 and 
 $\lambda$ from 0.1 to 0.7 in Scenario II. Except for the well-investigated combination,
the larger $\tan \beta$ and smaller $\lambda$ regime is also quite promising: $\Delta a_\mu$ 
can be increased  effectively due to the larger $\tan \beta$, and the SM-like Higgs boson ($H_2$) mass
can still be lifted by the pushing up effect because of the relatively large $A_{\lambda}$.

 \item Higgs Sector
 
 For Higgs sector, we present the plots for $R_{\gamma \gamma}-R_{VV}$ and $R_{Vbb}-R_{\tau \tau}$, 
as well as the masses of $m_{H_1}$, $m_{H_2}$, $m_{A_1}$, and $m_{H^{\pm}}$ ($H_3$, $A_2$ and $H^{\pm}$
are almost mass degenerate since they are all $H_d$-like.). In the
$m_{H_1}$-$m_{H_2}$ planes, the horizontal lines correspond to the cases in which the second lightest
CP-even neutral Higgs boson $H_2$ is SM-like and has a mass in the range [124, 127] GeV, while the vertical lines 
correspond to the cases of the lightest CP-even neutral Higgs boson $H_1$ being SM-like. At the intersection, 
these two Higgs bosons will all contribute to the LHC signals together in 
the manner presented in the previous subsection. From the 1$\sigma$ region of $\chi^2$ analyses, 
we obtain that the second lightest CP-even Higgs boson is more likely to be the SM-like Higgs boson
discovered at the LHC, as pointed out in many previous literatures. The only exception is a small 
region in Scenario III where $H_1$ is the SM-like Higgs boson and
$H_2$ is still lighter than about 180 GeV. Thus, in the most favourable
parameter space, we have 
another lighter Higgs boson with mass less than 125 GeV, which has reduced couplings with the SM particles. 
The corresponding signal strengths for dominant decay channels of the Higgs boson
discovered at the LHC are presented 
in the $R_{\gamma \gamma}-R_{VV}$ and $R_{Vbb}-R_{\tau \tau}$ planes. The 1$\sigma$ region generically predicts
that $R_{\gamma \gamma}$ ($R_{Vbb}/R_{\tau \tau}$) is a little bit larger (smaller) than 1.0.
  In the $m_{A_1}-m_{H^\pm}$ plane, 
we can see that $A_1$ is singlet-like and always lighter than 800 GeV, and the smaller values of $m_{A_1}$ 
are more favored. The masses of the $H_d$-like Higgs bosons have wide ranges from a few hundred GeV to several TeV,
and can be lighter than 1 TeV for small $\tan\beta  \sim 2$. However, when $\tan\beta$ 
increases, the masses of the $H_d$-like Higgs bosons will increase as well
since they are approximately proportional to $\tan\beta$, rendering 
heavier $A_2$, $H_3$, and $H^\pm$. The more specific discussions about large $\tan\beta$ regime will be given
in subsection~\ref{ltb}. It is obvious that the 1$\sigma$ regions are smaller in Scenario I compared to 
the Scenarios II and III since the requirement of the correct dark matter relic density imposes some constraints 
on the LSP components. The appropriate singlino fraction in the LSP neutralino is required  
since it is dominantly Higgsino like in Scenario I. We will comment on it in the following analyses about the LSP.


 \item Relevant Sparticle Masses
 
 As for sparticles, we present two panles, one ($m_{\tl g}-m_{\tl t_1}$) for colored sparticles and 
the other ($m_{\tl \chi_2^{\pm}}-m_{\tl \mu_L}$) for the representative sparticles in electroweak SUSY 
sector which invole in $\Delta a_{\mu}$. The narrow strips corresponding to 1$\sigma$ regions of $\chi^2$ analyses
in $m_{\tl g}-m_{\tl t_1}$ planes indicate that the colored sparticles such as light stop and
gluino could be either heavy 
about several TeV or relatively light $\lesssim$ 1 TeV. We observe that the light stop could 
be as light as 400 GeV in 1$\sigma$ regions. The light stop ($\lesssim 600$ GeV) regions could only 
be viable with small $\tan \beta$ and large $\lambda$ since the large tree-level contributions
to the SM-like Higgs boson mass are needed. When $\tan\beta$ increases, the larger light stop mass is required 
to compensate the reduction of the tree-level contribution to the SM-like Higgs boson mass. Since the LHC has 
given stringent constraint on the colored sparticle productions, we also show the naive bounds 
on the light stop and gluino masses for reference. Just keep it in mind that the light stop lighter than 
$\sim$ 600 GeV and gluino lighter than $\sim$ 1.3 TeV may be excluded with the LHC searches. 
Actually, we do not need to worry about this issue in the  EWSUSY since the following analyses of 
the LHC neutralino/chargino and slepton searches will push $\tan \beta$ to be large which corresponds to 
heavy squarks and gluino.
 
 In the $m_{\tl \chi^{\pm}_2}-m_{\tl \mu_L}$ plots, $\tl \chi_2^\pm$ and $\tl \mu_L$ are always as light as 
several hundred GeV, which is required by the EWSUSY. Because Higgsinos are always lighter than about
$ 300$ GeV by the virtue of small $\mu_{\rm eff}$, $\tl \chi_2^\pm$ is  Wino like in most of the 
parameter space. 
When $\tl \chi_2^\pm$ is very light, Wino and Higgsino would have large mixing. Given a small 
$\tan \beta$, very light smuon and Wino are prefered by $\Delta a_{\mu}$. Such light charginos 
and sleptons may be excluded from the LHC SUSY searches~\cite{lhcgau}, which will be discussed
 in the next Section.

 We do not present the neutralino sector here, since it is just trivial. Five neutralinos 
are all light within several hundred GeV. In the basis 
$\tilde \chi^0=(-i\tilde B, -i\tilde W, \tl H^0_d, \tl H^0_u, \tl S)$, 
the neutralino mass matrix is~\cite{Ellwanger:2009dp}
\begin{displaymath}
 \mathbf{M_{\tilde \chi^0}} =
 \left( \begin{array}{ccccc}
         M_1  & 0   & -\frac{g_1 v_d}{\sqrt{2}} & \frac{g_1 v_u}{\sqrt{2}} & 0 \\
              & M_2 & \frac{g_2 v_d}{\sqrt{2}} & -\frac{g_2 v_u}{\sqrt{2}} & 0 \\
              &     &      0            &  -\mu_{\rm eff} & -\lambda v_u \\
              &     &                   &  0          & -\lambda v_d \\
              &     &                   &             & 2\kappa s
        \end{array} \right)~.
\end{displaymath}
We have small $\mu_{\rm eff}(=\lambda s)$, which renders a lighter singlino due to
 $2\kappa s \sim 2 \frac{\kappa}{\lambda} \mu_{\rm eff}$. In other words, the singlino is 
lighter than about 200~GeV, and has large mixings with two
light Higgsinos. Since the Bino and Wino masses are free in principle, they could be 
either much heavier than $\sim 200-300$ GeV, or as light as $\sim 300$ GeV. When they 
are heavier, the typical neutralino order from light to heavy is 
$\tl H_{u/d}$, $\tl H_{d/u}$, $\tl S$, $\tl B$, $\tl W$. In most cases, the fisrt three states would have
large mixings, while the last two are dominated by Bino and Wino respectively. Thus, we have 
almost mass degenerate $\tl \chi_5^0$ and $\tl \chi_2^\pm$ being Wino-like. On the other hand, 
when five neutralinos are all very light, there exist large mixings among them. All these features 
would take effects in the collider searches for electroweak SUSY sector.

\item The Lightest Supersymmetric Particle (LSP)

\begin{figure}[htb!]
 \begin{center}
  \includegraphics[width=0.8\textwidth]{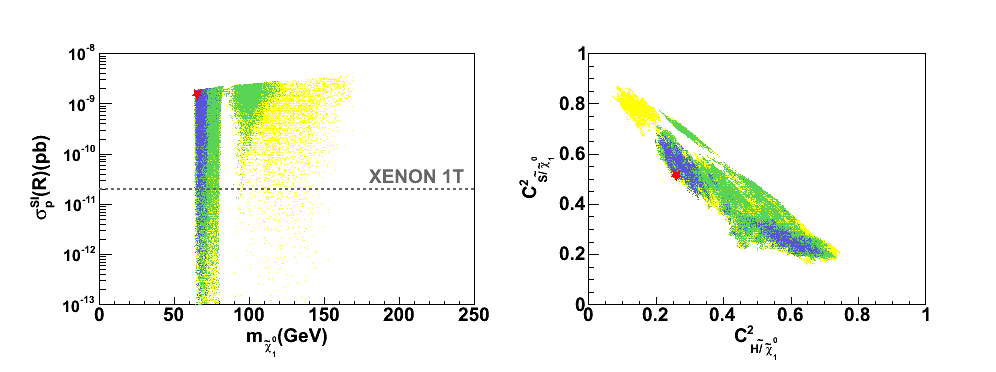}  \includegraphics[width=0.8\textwidth]{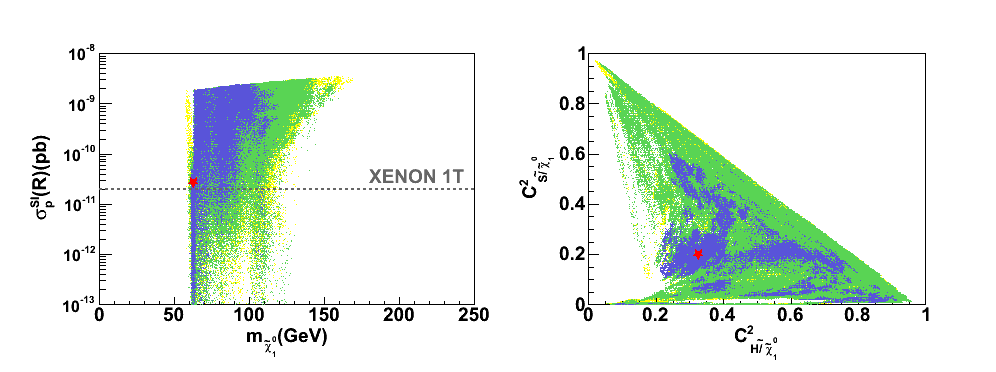}
  \caption{\label{lsp} The upper and down two plots are for Scenarios I and II, respectively.
In the left two plots, we present 
the rescaled spin-independent LSP neutralino-proton scattering cross section 
$\sigma_p^{SI}(R)=\sigma_p^{SI}\times \Omega h^2/0.11$ versus the LSP neutralino mass. 
The expected XENON1T sensitivity ($\sim 10^{-11}$pb) is shown as well. In the right two plots, 
we give the LSP Higgsino and singlino components in $C^2_{\tl H/\tl \chi^0_1}-C^2_{\tl S/\tl \chi^0_1}$ planes.}
 \end{center}
\end{figure}

We show the LSP neutralino relic density  for Scenarios I and II respectively 
in Figs.~\ref{s1} and \ref{s2}. It is obvious that the WMAP range for dark matter
 relic density  dominates the $\chi^2$ values in Scenario I. Just a narrow strip survives there 
with the LSP mass range $\sim [60,~80]$ GeV in 1$\sigma$ region of $\chi^2$ analyses. And in Scenario II, 
1$\sigma$ region maps to wider LSP mass range, from 60 GeV to 120 GeV,
and the LSP neutralino relic density could be very small due to its large Higgsino component. 
We can write the LSP neutralino mass eigenstate in terms of gauge eigenstates
\begin{equation}
 \tl \chi_1^0= C_{\tl B/\tl \chi_1^0}(-i\tilde B)+C_{\tl W/\tl \chi_1^0}(-i\tl W)+ C_{\tl H_d/\tl \chi_1^0}\tl H_d+C_{\tl H_u/\tl \chi_1^0}\tl H_u+ C_{\tl S/\tl \chi_1^0} \tl S~.
\end{equation}
Higgsinos have large annihilation cross sections into the SM particles, and then
the relic density for the Higgsino dominant LSP neutralino will be smaller than the observed
unless the LSP neutralino is a few TeV. Also,  Higgsinos have large scattering
cross sections with the SM particles. Thus, some singlino components are needed to obtain 
the correct dark matter relic density and satisfy the XENON100 experimental bound. We display the 
spin-independent LSP neutralino-proton scattering cross section and the LSP Higgsino and singlino
components  in Fig.~\ref{lsp}. Because of the large Higgsino component in the LSP, 
the XENON100 experiment 
would impose tight constraints on the LSP neutralino-proton scattering processes. One can see that 
the best-fit point in Scenario I is just below the current bound curve. We also show 
the expected XENON1T sensitivity ($\sim 10^{-11}$pb) in the plots. The majority of parameter space
could be detected at the XENON1T in the next a few years. The $C^2_{\tl H/\tl \chi^0_1}-C^2_{\tl S/\tl \chi^0_1}$ 
(with $C^2_{\tl H/\chi^0_1} \equiv C^2_{\tl H_u/\chi^0_1}+C^2_{\tl H_d/\chi^0_1}$)  plots differs 
in Scenarios I and II. In Scenario I, the LSP neutralinos are almost composed of Higgsinos and 
singlino only, while the moderate Wino and/or Bino components may invole in Scenario II. 
As expected, the moderate to large singlino component in the LSP is required in Scenario I.

In Scenario III due to RPV, the heavy sparticles will finally decay into the SM particles 
through the RPV superpotential terms. We present a plot of $m_{\tl \chi_1^0}-m_{\tl \tau_1}$ in Fig.~\ref{s3}.  
In the EWSUSY, the LSP could be $\tl \chi_1^0$, $\tl \tau_1$, or $\tl \nu_\tau$. The mass order 
among them would influence the cascade decays of the sparticles produced at the colliders, 
resulting in different signatures with the same heavy sparticle spectra. 

\end{itemize}


In summary, we have two different viable regimes: (1) the light stop and gluino are relatively light within
 1 TeV. And then the small $\tan \beta$ and large $\lambda$ are prefered by the 125 GeV 
SM-like Higgs boson, and somewhat lighter neutralinos/charginos and smuon/muon-sneutrino are required to make up 
 $\Delta a_{\mu}$; (2) The squarks and gluino are all heavy around several TeV. With the substantial 
radiative corrections to the SM-like Higgs boson mass from heavy stops, the relatively larger $\tan \beta$ is 
favored by $\Delta a_{\mu}$ confronted with the constraints from the LHC neutralino/chargino and slepton searches,
which will be discussed in the following. 

\begin{table}[ht]
 \begin{center}
 \scalebox{0.9}{
 \begin{tabular}{rl}
  \begin{tabular}{|c|c|c|c|}\hline
   $\chi^2_{min}$ Points: & I & II & III \\ \hline\hline
   \multicolumn{4}{|c|}{Param. at $M_{\rm GUT}$:(GeV)} \\ \hline
   $M_0$ & 1758 & 1576 &  1580\\ 
   $M_L$ & 7.1    & 21 &  47\\ 
   $M_E$ & 304    & 251 &  272\\
   $M_1$ & 348   & 256 &  343\\
   $M_2$ & 209    & 202 &  182\\ 
   $M_3$ & 556  & 336 &  585\\
   $A_0$ & -1955    & -2354 &  -2480\\
   $A_E$ & -7964   & -4801 &  -5569\\ 
   $A_{\lambda}$ & -618 & -854 &  -751\\ 
   $A_{\kappa}$ & -0.02 & -0.01 &  -0.02\\  \hline
   
   \multicolumn{4}{|c|}{Param. at $M_{\rm SUSY}$:}  \\ \hline
   $\lambda$ & 0.615 & 0.650 &0.582\\ 
   $\kappa$ &  0.082& 0.134 & 0.085\\ \
   $\tan\beta$ &1.94& 1.98 &1.97\\
   $\mu_{\rm eff}$ & 204 & 189 & 254\\   \hline \hline
   
   \multicolumn{4}{|c|}{Spectrum:(GeV)}\\  \hline
   $H_1$ & 88 & 111 & 99\\ 
   $H_2$ & \bf{125.2} & \bf{125.4} & \bf{124.8}\\
   $H_3$ & 490 & 448 & 616\\
   $A_1$ & 108 & 85 & 111\\ 
   $A_2$ & 494 & 450 & 618\\
   $H^{\pm}$ & 482 & 437 & 610\\    \hline
   $\tl \chi^0_1$ & 65 & 62 & 77\\ 
   $\tl \chi^0_2$ & 117 & 115 & 120\\ 
   $\tl \chi^0_3$ & 152 & 140 & 145\\ 
   $\tl \chi^0_4$ & -242 & -228 & -282\\
   $\tl \chi^0_5$ & 277 & 266 & 306\\ 
   $\tl \chi^{\pm}_1$ & 111 & 104 & 110\\
   $\tl \chi^{\pm}_2$ & 268 & 258 & 296\\
   $\tl g$ & 1360 & 873 &  1413\\   \hline
   

  \end{tabular}
  \begin{tabular}{|c|c|c|c|}\hline 
   $\chi^2_{min}$ Points: & I & II & III\\  \hline\hline
   $\tl \nu_{e/\mu}$ & 158 & 127 & 181\\
   $\tl \nu_{\tau}$ & 125 & 112 & 167\\  
   $\tl e_R/\tl \mu_R$ & 221 & 197 & 133\\ 
   $\tl e_L/\tl \mu_L$ & 168 & 140 & 190\\ 
   $\tl \tau_1$ & 93 & 108 & 64\\
   $\tl \tau_2$ & 203 & 190 & 190\\
   $\tl t_1$ & 859 & 472 & 750\\
   $\tl t_2$ & 1051 & 704 & 1021\\
   $\tl b_1$ & 996 & 632 & 970\\
   $\tl b_2$ & 2479 & 2119 & 2298\\
   $\tl u_R/\tl c_R$ & 2497 & 2129 & 2325\\ 
   $\tl u_L/\tl c_L$ & 1916 & 1571 & 1821\\ 
   $\tl d_R/\tl s_R$ & 2497 & 2133 & 2322\\
   $\tl d_L/\tl s_L$ & 1916 & 1572 & 1822\\    \hline \hline

   \multicolumn{4}{|c|}{Pheno.}  \\  \hline
   $R_{\gamma \gamma}$ & 1.32 & 1.28 & 1.36\\ 
   $R_{VV}$ & 1.02 & 1.02 &  0.97\\ 
   $R_{Vbb}$ & 0.90 & 0.80 &  0.82\\
   $R_{\tau \tau}$ & 0.94 & 0.85 & 0.85\\ 
   ${\rm BR}(b \rightarrow s\gamma) /10^{-4}$& 3.77 & 3.54 & 3.54\\ 
   ${\rm BR}(b \rightarrow \tau \nu)/10^{-4}$ & 1.32 & 1.32 & 1.32\\
   ${\rm BR}(B_s \rightarrow \mu^+ \mu^-) /10^{-9}$& 3.67 & 3.67 & 3.67\\ 
  
   $\Delta a_{\mu}/10^{-9}$ & 2.64 & 2.85 & 2.74\\ 
   $\Omega h^2$ & 0.1157 & $7\times10^{-5}$ &-- \\ 
   $\sigma^{SI}_p/10^{-9}pb$ & 1.6 & 46 &-- \\ \hline

   \end{tabular}
   
 \end{tabular} } 
 \caption{\label{bptable} The best $\chi^2$ fit points for three scenarios:
 $\chi^2_{min}/{\rm DOF}$=10.2/15,  $\chi^2_{min}/{\rm DOF}$=9.6/14,
and $\chi^2_{min}/{\rm DOF}$=9.2/14 for Scenarios I, II, and III, respectively.}
 \end{center}

\end{table}

\section{The Searches for the Electroweak SUSY Sector at the Colliders}
\subsection{ The EWSUSY Particle Spectra in the NMSSM Inspired by $\Delta a_{\mu}$}

Before the further phenomenological studies, we would like to summarize all the above theoretical and experimental 
results: 

\begin{itemize}

 \item In the NMSSM, to obtain a 125 GeV SM-like Higgs boson, we have small 
$\mu_{\rm eff} \sim 100-300$ GeV in the most favourable regions. 
Thus,  we have at least three light neutralinos ($\tl H_u,~\tl H_d,~\tl S$), and one 
light chargino ($\tl H^{\pm}$) around 100 GeV to 300 GeV.
 
 \item In Higgs sector,  the second lightest CP-even neutral
Higgs boson is SM-like in $1\sigma$ regions of  $\chi^2$ analyses
in Scenarios I and II, and in almost all the $1\sigma$ regions in Scenario III. The point is that
the diagonalization of Higgs boson mass matrix would push up the SM-like Higgs boson mass. 
With the lightest CP-even and CP-odd Higgs bosons being singlet-like, 
we have three light Higgs bosons $H_1$, $H_2$, and $A_1$. 
As for the other three $H_d$-like Higgs bosons, they would be either light within 1 TeV in the small $\tan \beta$ 
and large $\lambda$ regime, or heavy about several TeV in large $\tan \beta$ regime which will be argued to 
be more interesting in subsection~\ref{ltb} from the LHC SUSY searches.
 
 \item The EWSUSY motivated by $\Delta a_{\mu}$ predicts the light neutralinos, charginos, and
sleptons (either left-handed, or 
right-handed, or both of them), which form the complete light electroweak SUSY sector 
 imposed by the general NMSSM. 
 
 \item In principle, the squarks and gluino could be either light or heavy, which depend on the specific input parameters.
We will show that they are indeed heavy from the LHC SUSY searches, as predicted from the EWSUSY.

\end{itemize}

As one can see from the last Section, the light electroweak SUSY sector is definitely needed for the EWSUSY, while 
light stop and gluino in the NMSSM could be somewhat lighter than those in the MSSM due to the extra contributions 
to the SM-like Higgs boson mass from the tree-level F-term and pushing up effect. However,  light stop 
and gluino are strongly disfavored when the LHC SUSY search results are taken into account. 
Thus, we come back to our original EWSUSY picture: the squarks and/or gluino are heavy about several TeV, and 
out of the current LHC reach. After the following dedicated LHC SUSY search studies, we will comfirm that
 this assumption is indeed valid. 
Thus, we will concentrate on the searches for electroweak SUSY sector: neutralinos, charginos, and sleptons, 
with the colored sparticles totally decoupled.

The sparticles in electroweak SUSY sector have much smaller production cross sections at the LHC than squarks/gluino,
and they are only mainly pair produced in electroweak processes through $s$ channels via electroweak gauge boson
exchanges. For Wino with mass around 200--500 GeV, the Wino-like neutralino-chargino pair production 
cross section varies from $\sim 0.2$ pb to $\sim 0.002$ pb at the LHC with
${\sqrt s}=7$ TeV, while the slepton pair production has about $ 50$ 
times smaller cross section. The corresponding cross sections will increase by several times at the LHC-14.
 The ATLAS and CMS Collaborations have performed the searches for the pair productions of 
neutralinos/charginos and sleptons~\cite{lhcgau}. They focused on the Wino-like chargino-neutralino 
$\tl \chi_2^0 \tl \chi_1^\pm$ pair production and the slepton 
$\tl l \tl l^*$ pair production with Bino-like $\tl \chi_1^0$. Results are interpretated in simplified models. 
The specific decay chains are assumed for Winos: $\tl \chi_2^0 \to l \tl l \to l l \tl \chi_1^0$ and 
$\tl \chi_1^\pm \to l \tl \nu \to l \nu \tl \chi_1^0$($l=e,\mu, \tau$) for sleptons lighter than Winos; 
and $\tl \chi_2^0  \to Z^{(*)} \tl \chi_1^0$ and  $\tl \chi_1^\pm \to W^{\pm(*)} \tl \chi_1^0$ for heavy sleptons. 
The signatures of tri-leptons or same-sign di-leptons with missing energy would be produced in the cascade decays. 
The first case would produce quite a few leptons, while less leptons are produced and large SM backgrounds involve 
in the second case. $\tau$-enriched scenario is also considered, assuming the light right-handed sleptons and 
$\tl \chi_1^\pm$ decays through its Higgsino component. In these simplified topologies, Winos with mass less than
$ 600~$GeV have been excluded at 95\% C.L. for the lighter sleptons and the LSP lighter than 200 GeV. The mass 
upper limit is reduced to $\sim 320$ GeV in the absence of light sleptons. 
In addition, chargino and slepton pair productions are explored in the opposite-sign 
lepton-pair search channel. Interpreted in simplified models, the chargino mass can be explored up 
to $\sim$ 400 GeV, and the left-handed slepton masses are explored up to $\sim$ 200 GeV.

However, these bounds on sparticle masses can not be applied directly to the concrete models since
the productions and decays are much more complicated. First of all,  we need to 
consider the whole light electroweak SUSY sector, such as Higgsinos, sleptons, Bino, and Winos. So 
 we have more signal sources. Second, the decay chains would become longer, and various
 decay chains would entangle with each other. Generally speaking, heavy neutralinos and chargino would decay
 into lighter ones accompanied with electroweak gauge bosons or light Higgs bosons. With sleptons lighter than 
them, the slepton channels would be dominant decay modes and give rise to rich leptons in final states. 
Taking the best-fit point of Scenario I as an example, we have all the neutralinos and charginos lighter 
than $300$ GeV, and all the sleptons lighter than $\sim 200$ GeV. The neutralinos and charginos are all 
well mixed. With a quite light $\tl \tau_1$ (93 GeV), $\tl \chi_1^\pm$, $\tl \chi_2^0$, and $\tl \chi_3^0$ will all 
decay into $\tl \tau_1$ with $\sim$ 100\% branching ratios
 through their Higgsino components. However, the small mass splitting 
between $\tl \tau_1$ and the LSP, which leads to soft decay products, would block majority of the signals. 
$\tl \chi_4^0$, $\tl \chi_5^0$, and $\tl \chi_2^\pm$ are heavier and have enough phase space to decay to 
the lighter sleptons and on-shell gauge bosons, which will give rise to rich leptons in final states. 
$\tl \chi_2^\pm$ will decay into the light left-handed sleptons (with an exception of $\tl \tau_R$ since 
Higgsino component in $\tl \chi_2^\pm$ would decay into lighter $\tl \tau_R$), the lighter chargino or neutralinos,
and the $\tl \chi_1^\pm$ and light Higgses with branching ratios  40\%, 45\%, and 11\%, respectively.
$\tl \chi_5^0$ will decay into $\tl \chi_1^\pm W^\mp$ and sleptons respectively with 
branching ratios  32\% and 40\%, and decay into $\tl \chi_i^0 Z$ and $\tl \chi_i^0 H/A$
with small branching ratios. The longer cascade decay chains would give rise to more, although softer, leptons. 
The significant signature would be multi-leptons (including $\tau$). The more detailed features in the search for 
electroweak SUSY sector will be discussed in subsection IV D. In next subsection, we will apply 
the current LHC SUSY search results to our sampled points. Because the dominant signals come from the decays of 
heavier neutralinos/chargino and then the masses of left-handed sleptons would modify the decay modes 
drastically, we will interpret our results in terms of $m_{\tl \chi_2^\pm}$ (Wino mass typically) and $m_{\tl \mu_L}$,
  which are also the most important particles in SUSY contribution to $\Delta a_\mu$ in our study. 

\subsection{The LHC SUSY Search Bounds}

In this subsection, we apply the LHC neutralino/chargino and slepton search results to check our sampling points. 
The 8~TeV-9.2~fb$^{-1}$ results from the CMS Collaboration and the 8~TeV-13~fb$^{-1}$ results from 
the ATLAS Collaboration are employed. We consider the following analysis procedure: for each point, 
we generate events via Monte Carlo tools, and then apply the experimental selections and 
cuts to them. After we obtain the effective contribution in every signal region ($N_{\rm sig.}$),
we compare the model prediction  with the  95\% C.L. upper limit ($N_{\rm U.L.}$) provided by the experimental 
Collaborations. If any new contribution is larger than the limit ($N_{\rm sig.}/N_{\rm U.L.}>1$),  this point or
model is excluded. 

MadGraph5~\cite{Alwall:2011uj} is used to generate parton-level events and Pythia6.4~\cite{Sjostrand:2006za} 
is employed to perform parton shower and hadronization. Then events are passed to PGS4~\cite{pgs4} for 
detector simulation. With all the squarks and gluino so heavy, we only generate the processes 
for $\tilde \chi \tilde \chi$ and $\tilde l \tilde l$ pair productions, where $\tilde \chi$ includes 
all the neutralinos/charginos and $\tilde l$ represents all the sleptons.  Instead of the
complete calculations for every model points, a NLO k-factor of 1.2~\cite{Choi:2004zx} is taken for simplicity. 
Because the LSP would influence the sparticle cascade decay final states a lot, we will study 
 three scenarios separately. For Scenario III with RPV, Pythia8 and Delphes~\cite{Ovyn:2009tx} 
are employed instead of Pythia6.4 and PGS4.
  
\begin{figure}[htb]
 \begin{center}
  \includegraphics[width=0.48\columnwidth]{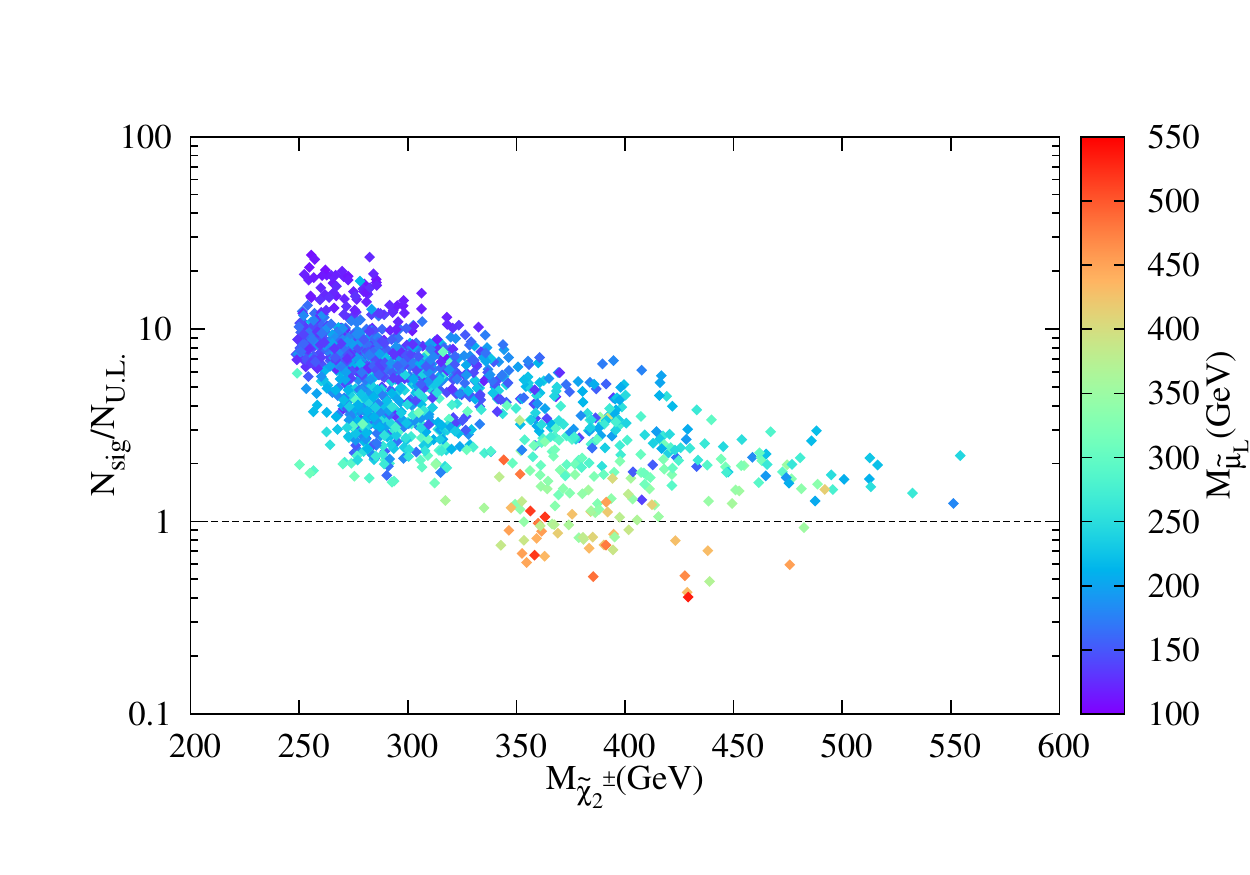}
  \includegraphics[width=0.48\columnwidth]{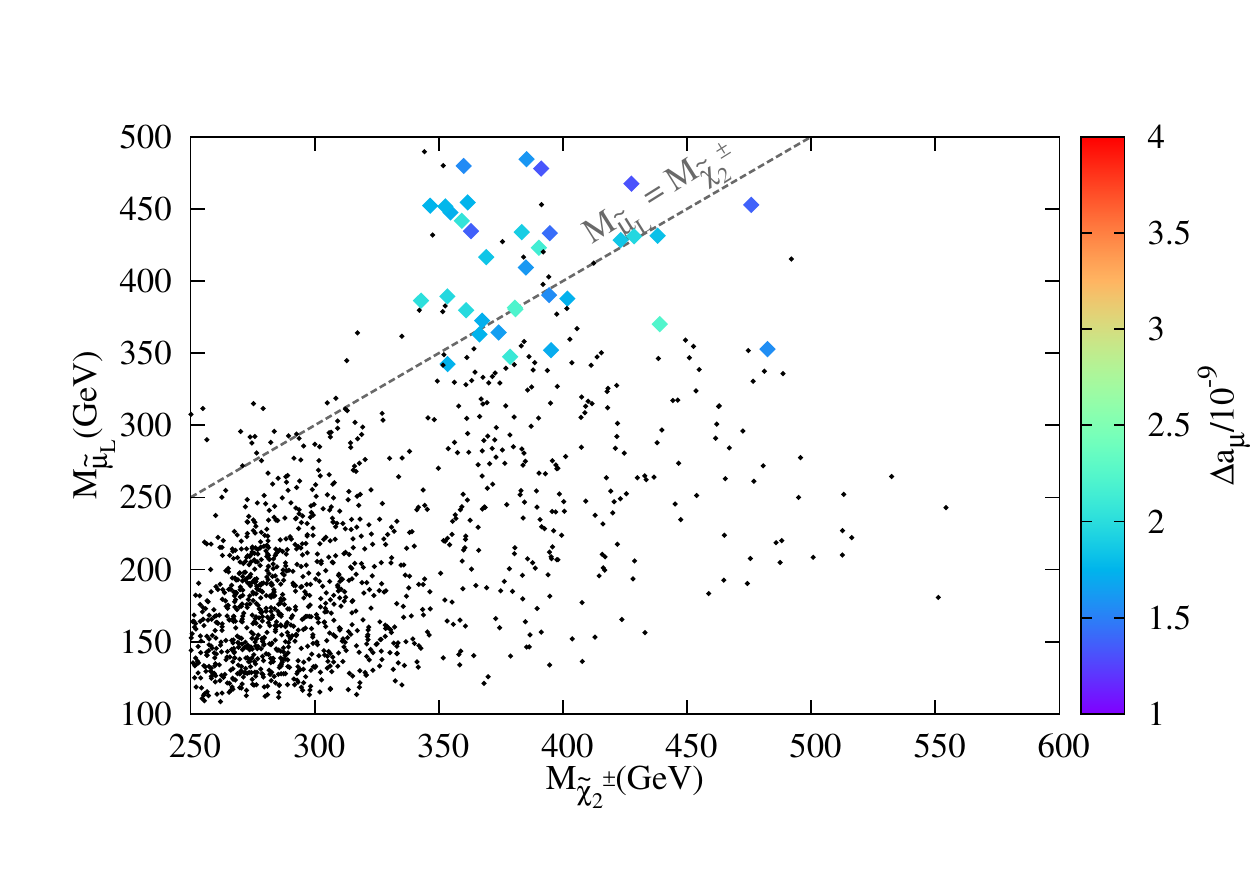}
  \caption{\label{lhc_s1}\textbf{Left}: The signal to U.L. ratio ($N_{\rm sig.}/N_{\rm U.L.}$) versus $M_{\tilde \chi_2^{\pm}}$
for all the sampled points in Scenario I. \textbf{Right}: $M_{\tilde \chi_2^{\pm}}-M_{\tilde \mu_L}$ plane. 
The dead points are in black, and the survived ones are marked with color suggesting values of $\Delta a_\mu$.}
 \end{center}
\end{figure}

For Scenario I, we have checked $\sim$ 1300 points which have $\Delta a_{\mu}$ in 2$\sigma$ range. 
Results are shown in Fig.~\ref{lhc_s1}. In the left plot, we present the signal to upper limit ratio
($N_{\rm sig.}/N_{\rm U.L.}$) respect to the $\tilde \chi_2^{\pm}$ mass, with $m_{\tl \mu_L}$ coded in color. 
One can see that most points have been excluded by current LHC SUSY searches. No points with $\tl \chi_2^\pm$ 
lighter than $\sim 350$ GeV survive here. And all the survived points  have relatively heavier 
$\tl \mu_L$ $\gtrsim 350$ GeV. In order to make it clear, we display the survived points 
in the $ m_{\tilde \chi_2^{\pm}}-m_{\tilde \mu_L}$ plane with colored markers for $\Delta a_{\mu}$ in the right plot, 
while coloring the excluded ones in black. One can see the values of $\Delta a_{\mu}$ through maker's colors.  
 The survived points mainly distribute in the small region with $M_{\tilde \chi_2^{\pm}} \lesssim M_{\tilde \mu_L}$
 or $\tl \chi_2^\pm$ and $\tl \mu_L$ are very mass degenerate. In this region, because sleptons are generally 
heavier than neutralinos and charginos, the pair-produced neutralinos and charginos will decay to lighter 
neutralinos/chargino and electroweak gauge bosons, corresponding to the simplified model with 
chargino and second neutralino decaying to the LSP through on shell or off shell W/Z bosons. The difference
is that the cascade decay chains are longer here, resulting more on-shell or off-shell gauge bosons in 
final states. Because the masses of Wino and sleptons are pushed up by the LHC SUSY search results, we obtain 
large $\tan \beta$ ($\gtrsim 11$) enforced by $\Delta a_\mu$ for survived points. Now with such large $\tan \beta$, 
the squark and gluino masses are automatically pushed up to several TeV in requirement of the 125 GeV
SM-like Higgs boson mass. Three survived benchmark points are given in Table~\ref{bptable1}, 
with branching fractions of dominant decay modes for relevant sparticles displayed as well. 
Because it is impossible to list all the decay modes for every sparticle, we just present some common  and 
important ones. Thus, some decay modes, which would be dominant in specific cases, may be missing in our table. 
The complete information could also be obtaind by simple physical estimation.
 
\begin{figure}[htb]
 \begin{center}
  \includegraphics[width=0.48\columnwidth]{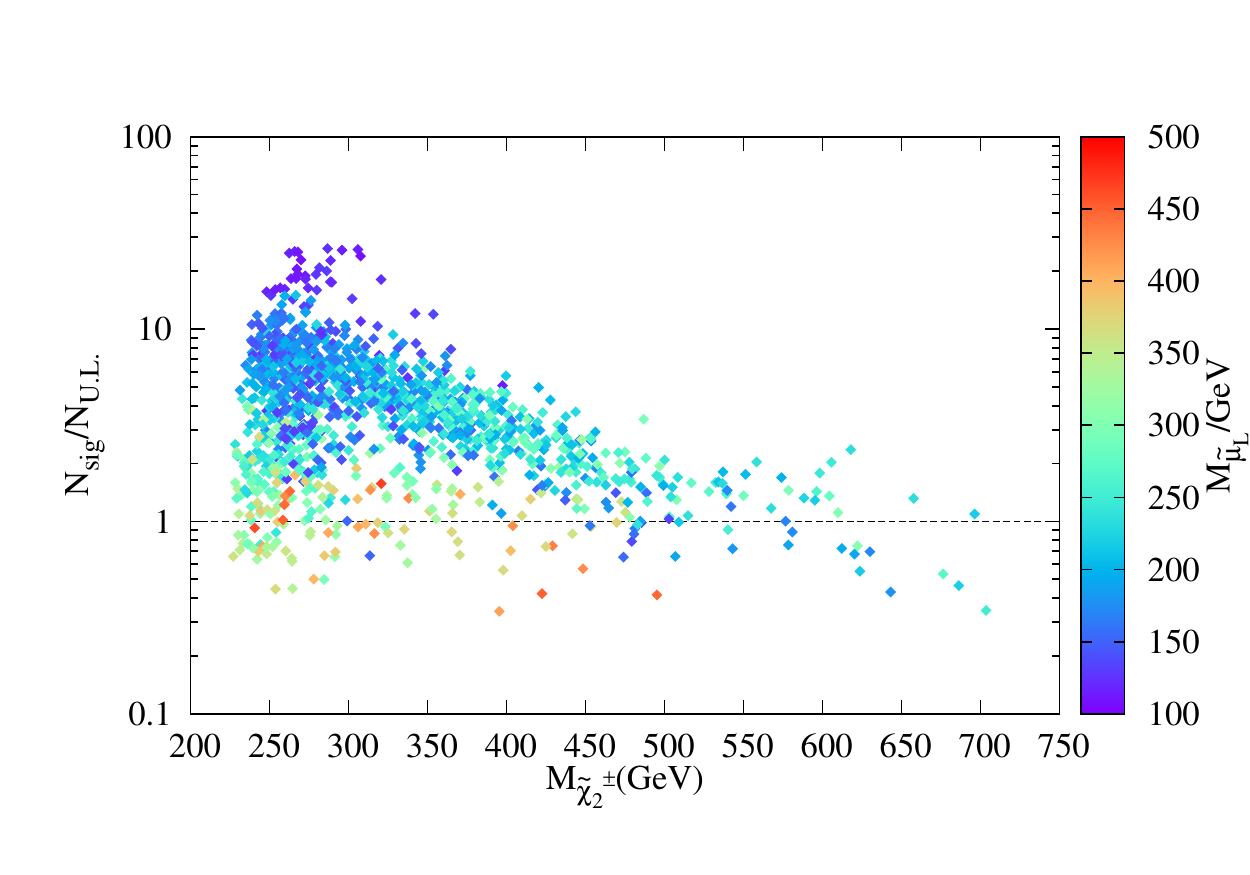}
  \includegraphics[width=0.48\columnwidth]{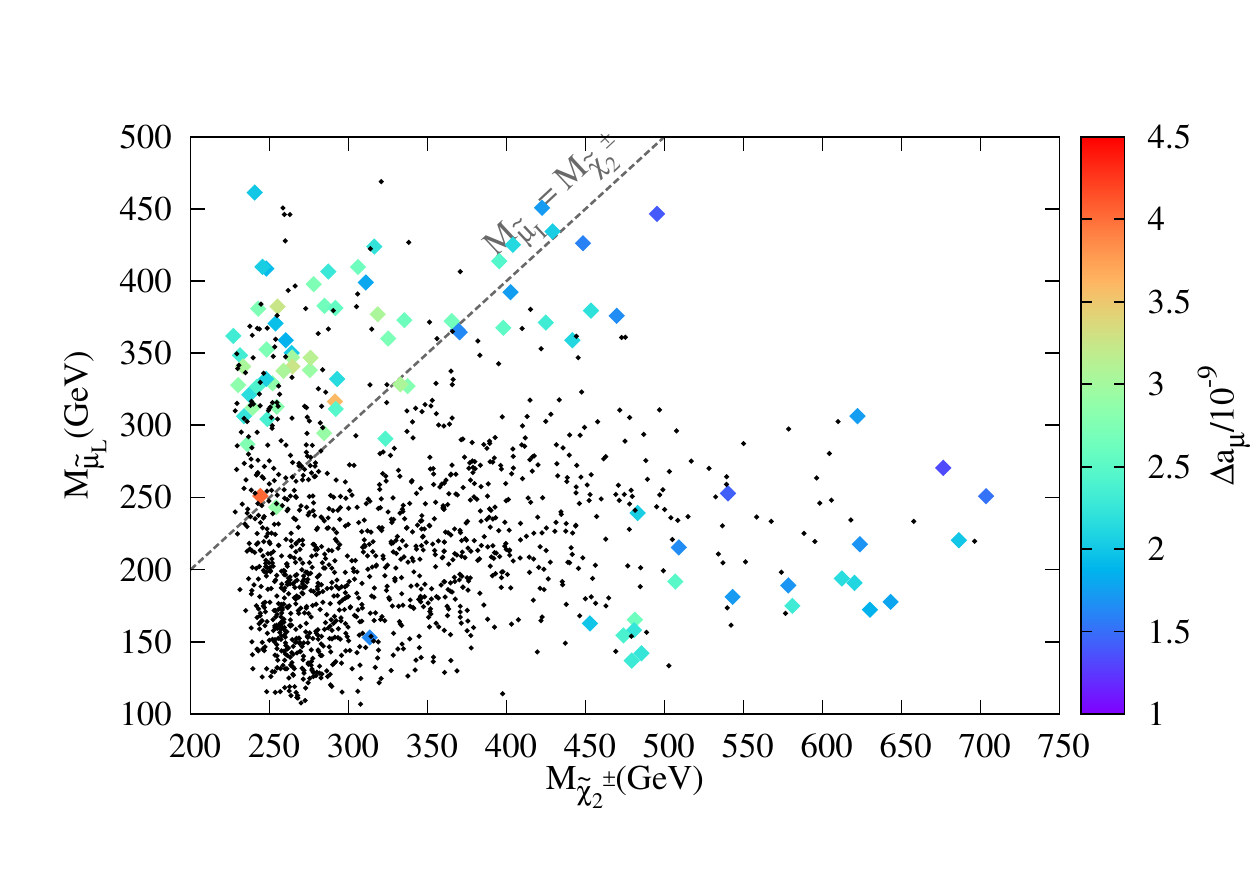}
  \caption{\label{lhc_s2}\textbf{Left}: The signal to U.L. ratio ($N_{\rm sig.}/N_{\rm U.L.}$) versus $ M_{\tilde \chi_2^{\pm}}$
for all the sampled points in Scenario II. \textbf{Right}: $ M_{\tilde \chi_2^{\pm}}-M_{\tilde \mu_L}$ plane. 
The dead points are in black, and the survived ones are marked with color suggesting values of $\Delta a_\mu$.}
 \end{center}
\end{figure}

As for Scenario II, the LHC SUSY search constraints  would be applied in the similar way as in Scenario I due to 
the similar particle spectra in light of the SM-like 
Higgs boson mass and $\Delta a_{\mu}$. We have also checked $\sim$ 1300 
sampled points with $\Delta a_\mu$ in 2$\sigma$ range
in Scenario II, and present the results in Fig.~\ref{lhc_s2}. We can see that there exist two 
main viable regions in the $m_{\tilde \chi_2^{\pm}}-m_{\tilde \mu_L}$ plane. One corresponds to the points with 
$\tilde \chi_2^{\pm}$ heavier than $\tl \mu_L$, and in this region the $\tilde \chi_2^{\pm}$ mass  is pushed up to 
about 450 GeV since the decay chains mediated by the light sleptons have large signal significances in the 
tri-lepton or same-sign di-lepton search channels. The other region locates at the up-left corner
with $M_{\tilde \chi_2^{\pm}} \lesssim M_{\tilde \mu_L}$, 
 and no definite bound on $m_{\tl \chi_2^\pm}$ can be obtained. So, unlike the Scenario I,
the very light $\tl \chi_2^\pm$ ($\sim 230$ GeV) 
are still available in this region. Similar to  Scenario I, almost all the
viable points have larger $\tan \beta$ ($\gtrsim 8$) and heavy 
squarks/gluino. We also give three survived benchmark points 
within  2$\sigma$  range of $\chi^2$ analyses
in Table~\ref{bptable2}.

Also, we find an interesting exception in Scenario II. One single point with $\tan \beta = 2.5$ survives. 
From the above discussion,  $\mu_{\rm eff}$ is larger than $M_2$ for this point.
 The LHC signals mainly come from heavier Higgsinos decays. 
However, Higgsinos have relatively smaller production cross sections, resulting in slightly reduced signals. 
Strictly speaking, we should present this point with $m_{\tl \chi_1^\pm}$ instead of $m_{\tl \chi_2^\pm}$
since $\tl \chi_1^\pm$ is Wino-like here. Thus, this point  just corresponds to the 
up-left region in the $m_{\tilde \chi_2^{\pm}}-m_{\tilde \mu_L}$ plot of Fig.~\ref{lhc_s2}.
For reference, we present this special point as Point IV in 
Table~\ref{bptable2}.



However, in Scenario III, the LHC phenomenology is somewhat different due to $R$-parity violation.
The LSP is not stable, and then can be the lightest neutralino, light stau, or tau-snetrino~\cite{Allanach:2006st}.
Also, there is no  missing energy for sparticle productions and decays at colliders.
The standard RPV superpotential in the NMSSM with $Z_3$ symmetry is
\begin{equation}
 W_{\sl R}=  \lambda^{\mu}_i S H_u L_i + \frac{1}{2} \lambda_{ijk} L_i L_j E^c_k +
\lambda'_{ijk} L_i Q_j D^c_k + \frac{1}{2} \lambda''_{ijk} U^c_i D^c_j D^c_k ~.~
\end{equation}
For simplicity, we do not consider $\lambda^{\mu}_i$ here, and
 take the other RPV trilinear couplings to be smaller than about $ 10^{-3}$ which will not change the
sparticle mass spectra. Thus, the sparticles will be dominantely produced in pairs 
(The RPV couplings are too small to render the resonant sparticle productions.). 
All the cascade decay chains will end in the SM particles due to
the RPV superpotential terms. The RPV coulings $\lambda$, $\lambda'$, and $\lambda''$ respectively correspond to 
the leptonic, semileptonic and hadronic decays, giving different final states.

We take only one RPV coupling (say, $\lambda_{121}$, $\lambda'_{311}$, or $\lambda''_{212}$) 
to be non-zero at one time. Turning on $\lambda$ will bring too many lepton signals 
which are strongly constrained. Thus, we will just consider the non-zero $\lambda'$ and $\lambda''$ here. 
For simplicity, we explore the RPV effects with several benchmark points, instead of performimg 
a complete scan over parameter space as in the first two Scenarios. 
With $\lambda'_{311}\neq0$, the $\tl \chi_1^0$-LSP will decay to one $\tau$ lepton and 
two jets ($\tl \chi_1^0 \to \tau^- u \bar d/\tau^+ \bar u d $), which gives an extra lepton 
while eliminates the original missing energy. Also, the $\tl \tau_1$-LSP would decay 
directly into two jets ($\tl \tau_1 \to \bar u d$).
We find that this extra lepton from LQD operator will make things worse, although no original 
missing energy present. Interestingly, the hadronic operator UDD relaxes the LHC constraints 
due to the missing energy suppression and no extra lepton in the final states. 
With $\lambda''_{212}\neq0$, the $\tl \chi_1^0$-LSP decays into three jets
($\tl \chi_1^0 \to  c d s/\bar c \bar d \bar s$). In this decay mode, the signal to U.L. ratio 
could be reduced by several times compared to the RPC case. By naively estimation, 
the $\tl \chi_2^\pm$ mass could be $\sim 50-100$ GeV lighter than those in Scenarios I and II. 
However, when the LSP is light stau or tau-sneutrino rather than neutralino, 
the rich $\tau$ signals from the Higgsino decays and then the LSP 
 light stau or sneutrino decays would give very remarkable signal
significances. And  the UDD operator would produce more $\tau$s than the LQD operator for 
the light stau LSP. So the viable Wino mass range gets back to the same level as in 
Scenarios I and II. Notice that these results are just for heuristic discussions, 
the systematic analysis is necessary if one is serious about these RPV effects.

By the way, we check the three best-fit points in 
Table~\ref{bptable}, and find that they are all excluded as expected from the
above discussions.

\subsection{The SUSY Models Consistent with $\Delta a_{\mu}$ and the LHC SUSY Searches}

We will discuss the general results in the SUSY models which are
consistent with $\Delta a_{\mu}$ and the LHC SUSY searches.

\subsubsection{Generic SUSY Models}

Although we consider the EWSUSY in the NMSSM to check the electroweak SUSY sector, 
the above results can be extended to general SUSY models which can
explain $\Delta a_\mu$. Although the small $\mu_{\rm eff}$ and 
an extra neutralino from singlino compared to the MSSM would 
modify the signals to some degree, the features of our results 
are reliable for more general SUSY models. To keep $\Delta a_\mu$ in the measured range, 
we need the light Higgsinos, Wino, and muon-sneutrino if
the chargino-sneutrino loops give dominant contributions.
These light sparticles are constrained by the LHC SUSY searches. Thus, 
the larger $\tan \beta$ is prefered, which is a general conclusion 
regardless of any specific SUSY model. If the left-handed sleptons
are lighter than Wino, the Wino with mass $\lesssim 450$ GeV   
are strongly disfavored. Otherwise, the Wino mass is generally not much 
constrained.
We have noticed that a general exploration in the MSSM has been made 
in Refs.~\cite{Endo:2013bba, Iwamoto:2013mya}. And our results are in agreement 
with theirs, although in different SUSY models.

Moreover, when $\mu$ is very large resulting in heavy Higgsinos, the
Bino-smuon contributions to $\Delta a_\mu$ would dominate by the virtue of 
large mixing (proportional to $\mu \tan \beta$) between the left-handed and right-handed smuons. 
Thus, only light Bino and sleptons are relevant sparticles as in Ref.~\cite{Ibe:2012qu}. And
the LHC SUSY search bounds can be almost evaded since the Bino production cross section
is smaller than Winos and Higgsinos.

\subsubsection{\label{ltb} The Moderate to Large $\tan \beta$ in the NMSSM}

The NMSSM with small  $\tan \beta$ has been studied extensively during the last year,
 inspired by the SM-like Higgs boson mass about 125 GeV and 
the possible excess of the Higgs decays to $\gamma \gamma$ channel observed at the LHC. 
Evidently, the small $\tan \beta$ regime is a salient feature of the NMSSM since
 the large tree-level contribution to Higgs boson mass can be realized for large $\lambda$. 
However, when we take all the known experimental results into account, 
the moderate to large  $\tan \beta$ in the NMSSM would be more interesting.
                                                                                               
In the constrained NMSSM, the large $\tan \beta$ is favored from a global analysis~\cite{Kowalska:2012gs} 
in light of the correct dark matter relic density, XENON100 dark matter search bound,
 and appropriate contributions to $(g_\mu-2)/2$, and regardless of which Higgs boson is the one observed 
at the LHC. However, because $\lambda$ is very small there, the results are similar to 
the MSSM. The mixing effects in Higgs sector are negligible  
as a consequence of such small $\lambda$~\cite{Kowalska:2012gs}. 

In this paper, we extend the above conclusion for the favorable larger $\tan \beta \sim \mathcal{O}(10)$ 
to a more general framework. With the EWSUSY, we have individual soft masses for sleptons, which would relax 
the tension between Higgs boson mass and $\Delta a_\mu$ even with small $\tan \beta \sim 2$. However, 
the LHC SUSY search results again prefer the relatively large $\tan \beta$.
In relatively large $\tan \beta$ regime, because the tree-level contribution from $\lambda S H_d H_u$ 
superpotential term to Higgs boson mass is negligible, the only new contribution comes from
the mixing effects between the doublet and singlet Higgs fields, with the SM-like Higgs boson being the 
second lightest one. So the whole contributions to Higgs boson mass are:
(1) the tree-level contribution $\sim 90$ GeV; (2) The radiative corrections from stop quarks $\sim 30$ GeV; 
 (3) The slight lift from the pushing up effect $\sim 5$ GeV.
The Higgs boson mass features change a little bit in this regime. 
For relatively large  $\tan \beta$, 
the $H_d$-like Higgs boson will be heavy at TeV scale (see below for details).
If we decouple it for simplicity,  the elements of Higgs boson mass matrix can be reduced to
\begin{align}
 M_{hh}^2 &\simeq m_Z^2 + \delta m_h^2 \sim 120 {\rm GeV}~, \\
 M_{ss}^2 &\simeq \lambda^2 v^2 \frac{A_\lambda}{\mu \tan \beta} + 4\left(\frac{\kappa}{\lambda}\right)^2 \mu^2 + \frac{\kappa}{\lambda} A_\kappa \mu~, \\
 M_{sh}^2 &\simeq 2 \lambda \mu v \left( 1- \frac{A_\lambda}{\mu \tan \beta} - \frac{2 \kappa}{\lambda \tan \beta} \right)~, 
\end{align}
where $\delta m_h^2$ denotes the radiative corrections, and the trilinear soft terms
$A_\lambda$ and $A_\kappa$ are the values at $M_{\rm SUSY}$. We also use $\mu$ instead of $\mu_{\rm eff}$ here 
for simplicity. For the original mass matrix, see Refs.~\cite{Ellwanger:2009dp, Kang:2012sy}.
 In order to increase the SM-like Higgs boson mass by several GeV, the 
singlet-like Higgs boson should be the lightest. Thus, the vaccum stability condition 
($ M_{sh}^4 \leq M_{ss}^2 M_{hh}^2 \lesssim (125 ~{\rm GeV})^4$) enforces 
a not so large $\lambda$ and 
$A_{\lambda} \sim \mu \tan \beta \sim \mathcal{O}(150 \times 10)~{\rm GeV}$. 
A small $\mu$ within several hundred GeV is also 
needed here to avoid the flipping in the mass order. 
Because $M_A^2 = \frac{2B_{\rm eff} \mu}{\sin 2\beta} \simeq \mu \tan \beta 
(A_\lambda+\frac{\kappa}{\lambda} \mu) \sim (\mu \tan \beta)^2$
for relatively large $\tan\beta$,
the $H_d$-like Higgs boson masses are estimated as $\mu \tan \beta$, which 
is around TeV scale. Therefore, we have two light CP-even neutral Higgs bosons, one light CP-odd Higgs boson, 
and the other three $H_d$-like Higgs fields ($H_3, A_2, H^\pm$) are heavy in this relatively 
large $\tan \beta$ and 
moderate $\lambda$ regime. Moreover, the heavy squarks and gluino
are needed to provide the desirable radiative corrections to Higgs boson mass.

\subsection{Prospects for the EWSUSY Searches at the LHC-14 and ILC}

After the LHC SUSY search constraints have been applied, 
the survived regions suggest that the EWSUSY might be just above the reach of 
the current detection capability. The light neutralinos/charginos and sleptons in the EWSUSY
 are very promising to be observed at the upcoming colliders such as the LHC-14 and ILC. 
Let us revisit the features of electroweak SUSY sector at first. We have 
$\tl \chi_1^0$, $\tl \chi_2^0$, and $\tl \chi_3^0$ lighter than $\sim 300$ GeV, 
which are mainly the mixtures of $\tl H_u$, $\tl H_d$ and $\tl S$. Also,
$\tl \chi_4^0$ and $\tl \chi_5^0$ respectively have dominant Bino and Wino components
in most cases. However, when all the neutralinos are within $\sim$300 GeV, there 
exist the large mixings among them, and no pure mass eigenstates remain. 
As for charginos, the Higgsino-like $\tl \chi_1^\pm$ are always lighter than 
about $ 200$ GeV, and $\tl \chi_2^\pm$ is Wino-like. Similar to the neutralino sector, 
when $\tl \chi_2^\pm$ is also very light, Higgsino and Wino will mix with each other as well.
Besides, the light left-handed smuon $\lesssim 500$ GeV are required by $\Delta a_\mu$, 
with the other left-handed sleptons and sneutrinos are also light for the sake of 
$SU(2)_L$ gauge symmetry and family universality. However, the right-handed sleptons 
are not neccessarily very light since their masses are controled by an independent 
input parameter $M_{\tl E}$. Their masses range from 100 GeV to 800 GeV in our scan results.

Because a bunch of sparticles may invole in, their mass order is very important for
 the sparticle decay chains. To be more concrete, we employ some benchmark masses  
$m_{\tl \chi_2^\pm}$, $m_{\tl \chi_1^\pm}$, $m_{\tl \mu_L}$, and $m_{\tl \tau_1}$ to organize the mass order,
where $m_{\tl \chi_2^\pm}$ and  $m_{\tl \chi_1^\pm}$ represent Wino and Higgsino masses, respectively. 
The most typical mass orders are listed as follows:
 (1)  $m_{\tl \chi_2^\pm} > m_{\tl \mu_L} > m_{\tl \tau_1}> m_{\tl \chi_1^\pm}$; 
(2) $m_{\tl \chi_2^\pm} > m_{\tl \mu_L} > m_{\tl \chi_1^\pm}> m_{\tl \tau_1}$; 
(3) $m_{\tl \mu_L}> m_{\tl \chi_2^\pm}   > m_{\tl \tau_1}>m_{\tl \chi_1^\pm}$; 
(4) $m_{\tl \mu_L}> m_{\tl \chi_2^\pm}   >m_{\tl \chi_1^\pm}> m_{\tl \tau_1}$. 
In these orders, the first two terms influence the Wino-like $\tl \chi_2^\pm$ and $\tl \chi_5^0$ decays, 
while the last two terms control the decays of the Higgsino-like $\tl \chi_1^\pm$ and $\tl \chi_{2/3}^0$. 
As for these sparticle productions, the cross sections for
the neutralino-chargino and chargino-chargino pair productions are the largest at the LHC. 
In the following, we will discuss the procuction processes separately:

\begin{itemize}
 \item Higgsinos: Higgsinos are very light and then have large production cross sections. 
They tend to decay into light stau and/or tau-snertrino if kinematically allowed, 
as in mass orders (2) and (4). So the rich $\tau$ signatures are noticable in this case. 
If light stau and tau-sneutrino are heavy as in mass orders (1) and (3), the Higgsino decays
will mediated by  virtual gauge bosons in most cases.  Unfortunately, the small mass difference between 
the produced Higgsinos and LSP blocks the visible final states, resulting in little observable signals. 
In general, we have $\Delta m \equiv m_{\tl \chi_i^{0/\pm}} - m_{\tl \chi_1^0} \sim 10-100$~GeV.
Ref.~\cite{Kraml:2008zr} discussed the signature of soft di-leptons at the LHC. Perhaps similar 
techniques could be applied to the Higgsino searches together with the Vector Boson Fusion (VBF)
production processes which could reduce the backgrounds drastically~\cite{Dutta:2012xe}. 
Actually, Higgsinos will serve as the more important roles in the
cascade decays of Bino and Winos than in direct production.
 
 \item Bino and Winos: Main production processes would be Wino pair productions. 
Bino has small production cross section, and mainly manifest itself in  Wino decays. 
With Higgsinos lighter than them, they will experience longer cascade decay chains. 
The Bino and Wino decays are highly dependent on the left-handed slepton mass. 
In mass orders (1) and (2), the left-handed sleptons are lighter and will take control of 
the decay chains, resulting in the golden tri-lepton signature 
in $\tl \chi^\pm \tl \chi^0$ pair-productions. Especially, in mass order (2), 
up to 7 leptons would emerge if Higgsinos all decay into $\tau$s. 
We list this long cascade decay chain here: 
$p p \to \tl \chi_2^\pm \tl \chi_5^0 \to (\tl l \nu)(\tl l l) \to (\tl \chi_4^0 l^\pm \nu)(\tl \chi_4^0 l^\pm l^\mp)$.
With Bino-like $\tl \chi_4^0$ mainly decaying into $\tl \tau_1 \tau$ ($\tl \tau_1$ is almost right-handed.), 
such decay chain ends in 
$(\tau^+ \tau^- l^\pm \nu \tl \chi_1^0)(\tau^+ \tau^- l^+ l^- \tl \chi_1^0)$. 
Ref.~\cite{Barger:2010aq} has discussed the many-lepton signatures from very long cascade decay chains 
in the NMSSM, although its particle content is a little bit different from here. On the other hand, 
in mass orders (3) and (4) with $m_{\tl \mu_L} > m_{\tl \chi_2^\pm}$ (From the subsection IV B, Winos
could be very light in these cases and then have very large production cross sections.),
Winos will generically decay into 
several gauge bosons on-shell or off-shell, depending on the mass splittings and specific mass order. 
Taking the decay chain
$p p \to \tl \chi_2^\pm \tl \chi_5^0 \to (\tl \chi_1^\pm Z)(\tl \chi_1^\pm W^\mp)$ as an example, 
it would end either in $(W^{\pm(*)} Z \tl \chi_1^0)(W^{\pm(*)} W^\mp \tl \chi_1^0)$ for mass order (4) 
or in 
$(\tl \tau_1 \nu_\tau Z)(\tl \tau_1 \nu_\tau W^\mp) \to (\tau \nu_\tau Z \tl \chi_1^0)(\tau^\pm \nu_\tau W^\pm \tl \chi_1^0)$ 
for mass order (3). Thus, the leptons ($\tau$ is very important in some cases) and gauge bosons 
in final states are typical signatures.
In addition to slepton and gauge boson decay modes, there would be several percent branching fraction into 
Higgs final states such as $\tl \chi_2^\pm \to \tl \chi_1^\pm (H/A)$ and 
$\tl \chi_5^0 \to \tl \chi_{1/2/3}^0 (H/A)$. The Higgs decay modes have been considered 
in Refs.~\cite{Baer:2012vr, Ghosh:2012mc}, which explored the $WH$ final states. As the concrete examples
for different decay patterns discussed above,  
one can refer to the benchmark points in Tables~\ref{bptable1} and \ref{bptable2}.
 
 \item Sleptons: When Winos are too heavy to explore, sleptons must be light enough to accomodate with
$\Delta a_\mu$. Thus, we could focus on the slepton pair productions instead. Light sleptons have 
large production cross sections at the ILC, and the opposite-sign di-leptons are typical signatures.

\end{itemize}

In short, the very interesting patterns would emerge in the searches for electroweak SUSY sector, which definitely
deserve further deep study. We just present several naive ideas here, and 
leave the dedicated analyses for a future work.

\section{Conclusion}

Taking into account all the available experimental constraints/results, especially
the 125 GeV SM-like Higgs boson at the LHC and the muon anomalous magnetic moment,
we have studied the EWSUSY in details in the NMSSM for three scenarios.
Using $\chi^2$ statistic test, we obtained the most favorable regions in the whole
parameter space. Moreover, we found that the LHC SUSY searches for neutralinos/charginos
and sleptons have already 
put considerable constraints on the electroweak SUSY sector.
And then the favored model parameter space and the resulting mass spectra
 are modified accordingly. After the systematic analyses, we are led to the
following conclusions:

\begin{itemize}

\item Unlike the previous studies in the NMSSM, we found that the moderdate to large 
$\tan \beta \sim \mathcal{O}(10)$ is prefered after the LHC SUSY search bounds are
taken into account. The relatively large $\tan \beta$ can be compatible
 with the muon anomalous magnetic moment and SM-like
Higgs boson mass simultaneously. Especially,
the SM-like Higgs boson is the second lightest CP-even neutral
Higgs boson, whose mass is lifted a little bit further by pushing up effect.

 \item The squarks and gluino are heavy around a few TeV and is out of the current 
LHC reach. The light electroweak SUSY sector lies on the brim of the current detection 
capability. In particular, the EWSUSY in the NMSSM can fit into all the current experimental data 
very well with 
$\chi^2/{\rm DOF} \sim 1$.  All the charginos, neutralinos, and sleptons are
  around several hundred GeV.
 
 \item The current LHC SUSY searches have put strong constraints on the light electroweak 
SUSY sector through the lepton final states. And the sparticle masses related
 to $\Delta a_\mu$ are constrained by these results as displayed in Figs.~\ref{lhc_s1} and \ref{lhc_s2}. 
Generically speaking, a Wino-like $\tl \chi_2^\pm$ lighter than $\sim 450$ GeV may be 
excluded when the left-handed sleptons are lighter than it.  However, with the left-handed sleptons 
heavier or nearly mass-degenerate with it, no definite constraint on the $\tl \chi_2^\pm$ mass
could be obtained. And a light $\tl \chi_2^\pm $ with mass around 230 GeV is still viable.

 \item The searches for electroweak SUSY sector of the EWSUSY in the
NMSSM is promising at the LHC-14 and ILC, 
and definitely deserve further dedicated analyses. The lepton final states are very important, and the
promising signatures include the multi-leptons (The number of leptons can be up to 7.), $\tau$ leptons, 
oppsite-sign di-leptons, and so on. In addition, the searches for Higgs bosons as final states would be
very promising if Winos are very light and then have large production cross sections.

\end{itemize}


\begin{table}[ht]
 \begin{center}
 \renewcommand{\arraystretch}{0.8}
 \addtolength{\tabcolsep}{3pt}
 \begin{tabular}{rl}
  \begin{tabular}{|c|c|c|c|}\hline
   Points: & I & II & III \\ \hline\hline
   \multicolumn{4}{|c|}{Param. at $M_{\rm GUT}$:(GeV)} \\ \hline
   $M_0$ &     2267        &   1746   &   2006    \\
   $M_L$ &     650         &   572    &   766     \\
   $M_E$ &     550         &   199    &   210     \\
   $M_1$ &     -348        &   -504   &   -360    \\
   $M_2$ &     376         &    442   &   353     \\
   $M_3$ &     -1435       &  -1924   &   -1429   \\
   $A_0$ &     60          &   121    &   121     \\
   $A_E$ &     -209        &  -573    &   -612    \\
   $A_{\lambda}$ &   4352  &  4625    &   4575    \\
   $A_{\kappa}$  &   836   &  421     &   418     \\   \hline
                                                  
    \multicolumn{4}{|c|}{Param. at $M_{\rm SUSY}$:}  \\ \hline
   $\lambda$	&    0.296  &  0.208  &  0.207   \\
   $\kappa$   	&    0.103  &  0.084  &  0.08  \\
   $\tan\beta$	&    16     &   19.8  &  19.2  \\
   $\mu_{\rm eff}$	&    162    &   129   &  143      \\   \hline  \hline
                                                 
   \multicolumn{4}{|c|}{Spectrum:(GeV)}\\ \hline 
   $H_1$  &    98      &  90  &  94       \\ 
   $H_2$  &      125.3     &  125  &  125     \\
   $H_3$  &      2883    &  2978  &  3151   \\
   $A_1$    &      82    &  68   &  73      \\
   $A_2$    &      2883    &  2978  &  3151   \\
   $H^{\pm}$&      2883     &  2979  &  3152   \\ \hline
   $\tl \chi^0_1$ &     100   &  92  &  99       \\
   $\tl \chi^0_2$ &     -127  &  -131 &  -123    \\
   $\tl \chi^0_3$ &     175   &  143  &  150     \\
   $\tl \chi^0_4$ &     -193  &  -217 &  -181    \\
   $\tl \chi^0_5$ &     378   &  439   &  359    \\
   $\tl \chi^{\pm}_1$ & 155   &  127  &  136     \\
   $\tl \chi^{\pm}_2$ & 379   &  439   &  359     \\
   $\tl g$ 	      & -3212 &  -4132    &  -3179    \\  
   $\tl t_1$ 		& 2763    &  3146 &  2646   \\
   $\tl q_{\rm min}$ 	& 3345    &  3810 &  3203          \\ \hline 

  \end{tabular}
  \begin{tabular}{|c|c|c|c|}\hline 
   Points: & I & II & III\\  \hline\hline
   $\tl \nu_{e/\mu}$   &   339    &  362  &  435    \\
   $\tl \nu_{\tau}$    &   200    &  108  &  249    \\
   $\tl e_R/\tl \mu_R$ &   961    &  724  &  921    \\
   $\tl e_L/\tl \mu_L$ &   348    &  370  &  442    \\
   $\tl \tau_1$        &   214    &  132  &  261    \\
   $\tl \tau_2$        &   879    &  532  &  770    \\
 \hline  \hline

   \multicolumn{4}{|c|}{Pheno.}  \\  \hline
   $R_{\gamma \gamma}$ 	&  1.1          & 1.04     &  1.0     \\ 
   $R_{VV}$            	&  0.99         & 1        &  0.96    \\ 
   $R_{Vbb}$ 		&  0.66        &  0.76     &  0.73             \\
   $R_{\tau \tau}$ 	&  0.67        &  0.77     &  0.74        \\
   ${\rm BR}(B \rightarrow X_s \gamma) /10^{-4}$	&    3.22    & 3.22     & 
3.21 \\ 
   ${\rm BR}(B \rightarrow \tau \nu_\tau)/10^{-4}$ 	&    1.32    & 1.31     & 
1.32 \\ 
   ${\rm BR}(B_s^0 \rightarrow \mu^+ \mu^-) /10^{-9}$& 3.68    & 3.68     & 
3.68\\ 
   $\Delta a_{\mu}/10^{-9}$ 	&   2.15                     & 2.32     &  2.1  
\\ 
   $\Omega h^2$	&   0.11                     & 0.103    &  0.1\\ 
   $\sigma^{SI}_p/10^{-9}pb$ 	&   2.2                      & 1.3   
&  2.8 \\ \hline \hline
    \multicolumn{4}{|c|}{BRs of Dominant Decay Modes}  \\  \hline  
   $\tl \chi_2^\pm \to \tl l_L \nu/\tl \nu l$ & 0.34 &0.51 &  0.22\\
   $\tl \chi_2^\pm \to \tl \chi_1^\pm Z$  & 0.16  & 0.12& 0.20\\   
   $\tl \chi_2^\pm \to \tl \chi_1^0 W^\pm$& 0.07  & 0.06& 0.09\\  
   $\tl \chi_2^\pm \to \tl \chi_2^0 W^\pm$& 0.05  & 0.11& 0.11\\  
   $\tl \chi_2^\pm \to \tl \chi_3^0 W^\pm$& 0.11  & 0.07& 0.13\\  
   $\tl \chi_2^\pm \to \tl \chi_1^\pm H_2$&  0.06 & 0.08 & 0.09\\\hline
   
   $\tl \chi_5^0 \to  \tl l_L l/\tl \nu \nu$&  0.35 & 0.52 & 0.22\\  
   $\tl \chi_5^0 \to \tl \chi_1^\pm W^\mp$&  0.35 & 0.26 & 0.46\\
   $\tl \chi_5^0 \to \tl \chi_2^0 Z$&  0.04 & 0.08 & 0.08\\ 
   $\tl \chi_2^\pm \to \tl \chi_1^0 H_2$& 0.02  & 0.02 & 0.03\\ \hline
   
   $\tl e_L/\tl \mu_L \to \tl \chi_1^\pm \nu$& 0.12 & 0.07& 0.07\\ 
   $\tl e_L/\tl \mu_L \to \tl \chi_1^0 e/\mu$& 0.09& 0.1& 0.05\\
   $\tl e_L/\tl \mu_L \to \tl \chi_2^0 e/\mu$& 0.61 & 0.27& 0.28\\ \hline
   
   $\tl \tau_1 \to \tl \chi_1^\pm \tau$ &0.08 & -- & 0.09\\
   $\tl \tau_1 \to \tl \chi_1^0 \tau$   &0.15 & 0.99& 0.13\\
   $\tl \tau_1 \to \tl \chi_2^0 \tau$   &0.68 & --& 0.51\\
\hline
\end{tabular}   
\end{tabular}
 \caption{\label{bptable1} The benchmark points satisfy the LHC SUSY search constraints
in Scenario I. Here, $\tl q_{\rm min}$ denotes the lightest squark in the first two generations.
 The kinematically forbidden or negligible decay modes  are presented in dashes.}       
\end{center}
\end{table}

\begin{table}[ht]
 \begin{center}
 \renewcommand{\arraystretch}{0.8}
  \addtolength{\tabcolsep}{3pt}
 \begin{tabular}{rl}
  \begin{tabular}{|c|c|c|c|c|}\hline
   Points: & I & II & III & IV \\ \hline\hline
   \multicolumn{5}{|c|}{Param. at $M_{\rm GUT}$:(GeV)} \\ \hline
   $M_0$ &   903        & 2699   & 390 & 1639\\
   $M_L$ &    28       &  12  &   24   & 30\\ 
   $M_E$ &    558       &  0.6  &  567 & 270\\ 
   $M_1$ &    -313       & 680   &  -517& 340\\
   $M_2$ &    278       &  762  &  163  & 221\\
   $M_3$ &    -1199       &558    &-1536& 520\\ 
   $A_0$ &    3453       & -3865   & 1818& -2873 \\ 
   $A_E$ &    -109       &  1.1  &   -80 & -2523\\ 
   $A_{\lambda}$ & 4115   & -0.7    &3862& -545 \\ 
   $A_{\kappa}$  & 224   &  -84  &  315  & -0.01\\  \hline
                                                
   \multicolumn{5}{|c|}{Param. at $M_{\rm SUSY}$:}  \\ \hline
   $\lambda$	&  0.19    &  0.34  &  0.31 & 0.613\\ 
   $\kappa$   	&  0.08    &  0.26  & 0.26  & 0.123\\ 
   $\tan\beta$	&  12.1    &  9.4  &    11.5& 2.48\\ 
   $\mu_{\rm eff}$	&  148    &  111  & 168 & 268\\ \hline   \hline
                
   \multicolumn{5}{|c|}{Spectrum:(GeV)}\\ \hline
   $H_1$  &   107     & 94     & 95  & 69 \\ 
   $H_2$  &   125.4     & 124.8  &  126.0 & 124.7  \\ \
   $H_3$  &   1996     & 1133   & 2085    & 752\\ 
   $A_1$    &   116     & 242  &  439     & 213\\ 
   $A_2$    &   1996     & 1132   & 2084  & 753 \\
   $H^{\pm}$&   1997     & 1133   & 2085  & 745 \\ \hline
   $\tl \chi^0_1$&      107       &87   &  112  &  97\\
   $\tl \chi^0_2$ &    -114  &-126  &  -167     & 142\\ 
   $\tl \chi^0_3$ &    150  &199  &  -229       & 164\\ 
   $\tl \chi^0_4$ &    -177  & 298  &  232      & -296\\ 
   $\tl \chi^0_5$ &    293  &  643  &  297      & 324\\ 
   $\tl \chi^{\pm}_1$ &130   &110   & 119       & 142\\ 
   $\tl \chi^{\pm}_2$ &293   &643    & 242      & 313\\ 
   $\tl g$ 	      & -2642  &   1421  &   -3298  & 1277 \\ 
      $\tl t_1$ 		&  1085   & 1641  &  1966 & 718 \\ 
   $\tl q_{\rm min}$ 	&  2403   & 2683  &  2850   & 1792 \\ \hline 

  \end{tabular}
  \begin{tabular}{|c|c|c|c|c|}\hline 
   Points: & I & II & III & IV\\  \hline\hline
   $\tl \nu_{e/\mu}$   &   323    & 160   &  318   &  138 \\ 
   $\tl \nu_{\tau}$    &   293    & 145   &  287   &  132 \\ 
   $\tl e_R/\tl \mu_R$ &   374    & 622   &  378   &  221\\ 
   $\tl e_L/\tl \mu_L$ &   332    & 178   &  327   &  153\\ 
   $\tl \tau_1$        &   301    & 164   &  296   &  143 \\ 
   $\tl \tau_2$        &   322    & 615   &  326   &  217\\ \hline \hline

   \multicolumn{5}{|c|}{Pheno.}  \\  \hline
   $R_{\gamma \gamma}$ 	&   1.02        & 1.29     &   1.11  & 1.11  \\ 
   $R_{VV}$            	&   0.97        & 1.18     &  1.02   & 1.02  \\
   $R_{Vbb}$ 		&    0.8      &  0.67     &  0.71    & 0.96   \\
   $R_{\tau \tau}$ 	&    0.8      &  0.68     &  0.72    & 0.96  \\ 
   ${\rm BR}(B \rightarrow X_s \gamma) /10^{-4}$	&   3.46     &  3.28    & 3.29
 & 3.33\\ 
   ${\rm BR}(B \rightarrow \tau \nu_\tau)/10^{-4}$ 	&    1.32    &  1.31    & 1.32
 & 1.32\\ 
   ${\rm BR}(B_s^0 \rightarrow \mu^+ \mu^-) /10^{-9}$&  3.68   & 3.66     & 3.68
 & 3.67\\ 
   $\Delta a_{\mu}/10^{-9}$ 	&     2.15          &  1.78    & 2.41   
 & 1.63\\ 
   $\Omega h^2$	&       0.02            &  0.03   & 0.002 & 0.02\\
   $\sigma^{SI}_p/10^{-9}pb$ 	&  11    &  0.9 &  39 & 11\\ \hline \hline
   
    \multicolumn{5}{|c|}{BRs of Dominant Decay Modes}  \\  \hline  
   $\tl \chi_2^\pm \to \tl l_L \nu/\tl \nu l$ & -- & 0.72& -- & 0.35\\
   $\tl \chi_2^\pm \to \tl \chi_1^\pm Z$  &  0.28 & 0.07& 0.33& 0.18\\   
   $\tl \chi_2^\pm \to \tl \chi_1^0 W^\pm$&  0.20 &0.06 & 0.57& 0.07\\  
   $\tl \chi_2^\pm \to \tl \chi_2^0 W^\pm$&  0.14 & 0.07& --  & 0.04\\  
   $\tl \chi_2^\pm \to \tl \chi_3^0 W^\pm$&  0.14 & --& --    & 0.22\\  
   $\tl \chi_2^\pm \to \tl \chi_1^\pm H_2$&  0.08  & 0.05 & 0.10& 0.06\\\hline
   
   $\tl \chi_5^0 \to  \tl l_L l/\tl \nu \nu$&  -- & 0.72 & -- & 0.35\\  
   $\tl \chi_5^0 \to \tl \chi_1^\pm W^\mp$&  0.70 & 0.15 & 0.48& 0.36\\
   $\tl \chi_5^0 \to \tl \chi_2^0 Z$&  0.10 &0.05 & 0.26 & 0.05\\ 
   $\tl \chi_2^\pm \to \tl \chi_1^0 H_2$& 0.03  & 0.03 & 0.11& 0.09\\ \hline
   
   $\tl e_L/\tl \mu_L \to \tl \chi_1^\pm \nu$& 0.19 &0.82 & 0.37 & 0.63\\ 
   $\tl e_L/\tl \mu_L \to \tl \chi_1^0 e/\mu$& 0.13&0.10 & 0.26  & 0.28\\
   $\tl e_L/\tl \mu_L \to \tl \chi_2^0 e/\mu$& 0.34 &0.09 & 0.07 & 0.09\\ \hline
   
   $\tl \tau_1 \to \tl \chi_1^\pm \tau$ &0.24 & 0.12& 0.43 & 0.02\\
   $\tl \tau_1 \to \tl \chi_1^0 \tau$   &0.15 & 0.63& 0.31 & 0.98\\
   $\tl \tau_1 \to \tl \chi_2^0 \tau$   &0.44 & 0.25& 0.08 & --\\

   \hline
\end{tabular}   
\end{tabular}
 \caption{\label{bptable2}The benchmark points satisfy the LHC SUSY search constraints
in Scenario II. Here, $\tl q_{\rm min}$ denotes the lightest squark in the first two generations.
 The kinematically forbidden or negligible decay modes  are presented in dashes.}                                               
 \end{center}
\end{table}


\begin{acknowledgments}

This research was supported in part
by the Natural Science Foundation of China
under grant numbers 10821504, 11075194, 11135003, 11275246, 
and by the DOE grant DE-FG03-95-Er-40917 (TL).

\end{acknowledgments}

\end{document}